\documentstyle[12pt, aaspp4, flushrt]{article}
\lefthead{Barmby \& Huchra}
\righthead{Hercules supercluster}
\begin{document}

\title{Kinematics of the Hercules Supercluster}
\author{Pauline Barmby}
\and
\author{John P. Huchra}
\affil{Harvard-Smithsonian Center for Astrophysics}
\authoraddr{60 Garden St., Cambridge, MA 02138} 
\abstract
The Hercules Supercluster consists of the Abell clusters 2147, 2151,
and 2152. Previous studies of the kinematics have been confounded by the 
difficulty of correctly assigning galaxies to the individual clusters, 
which are not well-separated. Our study has a total of 468 available velocities 
for galaxies in the region, 175 of them new. 414 galaxies are in the 
supercluster, about 
three times the number used in the previous supercluster study. We
verify the existence of the three individual 
clusters and compute their individual dynamical parameters. 
We investigate several techniques for assigning galaxy membership to 
clusters in this crowded field. We use the KMM
mixture-modeling algorithm to separate the galaxies into clusters; we find
that A2152 has a higher mean velocity than previous studies have reported. 
A2147 and A2152 also have lower velocity dispersions:  821$^{+68}_{-55}$ and 
715$^{+81}_{-61}$~km~s$^{-1}$. 
The assignment of galaxies to either A2152 or A2147 
requires velocity and position information.
We study the kinematics of the supercluster using the two-body formalism of 
Beers, Geller, and Huchra (1982) and conclude that A2147 and A2151 are probably
bound to each other, and that the supercluster as a whole may also be bound.  
The mass of the supercluster, if bound, is 
$(7.6\pm2.0)\times 10^{15} h^{-1} M_{\sun}$; with the supercluster 
luminosity, $(1.4\pm0.2) \times 10^{13} h^{-2} L_{\sun}$, this yields 
$\Omega \approx 0.34\pm0.1$. 

\section{Introduction}

Ostriker, Peebles, \& Yahil (1974) were among the first to suggest that
the mass-to-light ratios of spiral galaxies increase with increasing scale.
They noted that this trend appeared to continue to larger scales, indicating
that galaxies might provide the critical mass density for the universe.
Bahcall (1997) has suggested that the evidence now shows that $M/L$ of 
galaxies increases only up to a scale of 0.1-0.2 Mpc and then levels off, 
remaining roughly constant for groups 
and clusters of galaxies up to a scale of 1.5 Mpc. It is important to know 
whether this holds on much larger scales, e.\ g.\ superclusters. 
If so, it might imply that there is no additional dark matter on 
supercluster scales (of~$\sim 10$~Mpc). 
Since the constant value of $M/L_B \sim 300h$ implies a value for $\Omega$
of roughly 0.2, this would imply that
an $\Omega=1$ universe would have to have most of the matter outside groups,
clusters, and superclusters. 
Results for a few superclusters seem to imply that the necessary dark
matter does not exist on supercluster scales. 
Mass determinations for the Local Supercluster imply 
$\Omega\sim 0.3$ (Huchra 1988), and Postman, Geller, \& Huchra's (1988) 
mass for the Corona Borealis supercluster
gives $\Omega = 0.2\pm0.1$ (they also state that an increase in $M/L$ from 1 Mpc
to 10-20 Mpc scales is not required to account for the supercluster's
observed peculiar velocities.) The Hercules Supercluster is a nearby, 
well-studied supercluster; determining its mass is a useful step in
studying the properties of dark matter on large scales.

\begin{figure}
\plotone{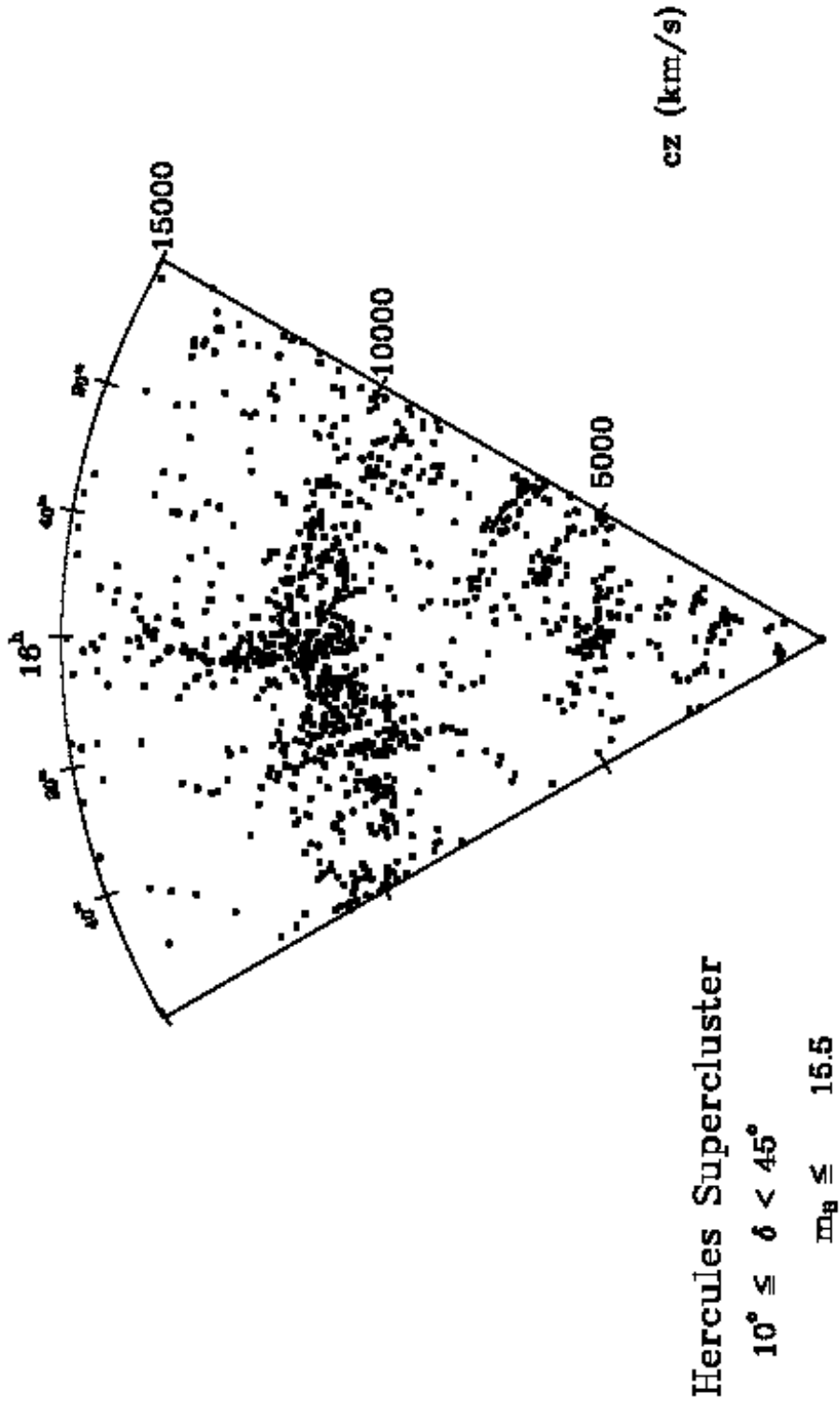}
\caption{Cone diagram for galaxies in the Hercules region. Center: Hercules supercluster. 
To the west are A2107, A2063, and A2052. \label{cone-wide}}
\end{figure}

The first person to point out the Hercules supercluster (\cite{sha34}) 
compared it to the Virgo cluster, and stated that ``[i]f it were not for its
great distance (perhaps thirty megaparsecs or more) the twin supergalaxy
in Hercules \dots would be equally interesting.'' 
The region (Figure~\ref{cone-wide}) contains two superclusters: 
the double cluster Abell~2197/99 ($\alpha \sim$ 16$^{\rm h}$, 
$\delta \sim 40 \arcdeg$, $v \sim $11000~km~s$^{-1}$), and the 
`Hercules supercluster', consisting of 
the rich cluster Abell~2151 (the `Hercules cluster')
and the double cluster Abell~2147/2152 (all at $\alpha \sim$~16$^{\rm h}$, 
$\delta \sim 17 \arcdeg$, $v \sim $11000 km~s$^{-1}$).
At a low density enhancement, all of these clusters, plus A2107, A2063, 
and A2052, can be considered part of a single supercluster 
(\cite{pos92}; \cite{bah84}).
We have obtained 175 new redshifts in a 22.5 square degree area 
which includes the Hercules supercluster; we combine these
with redshifts from the literature in a study of the kinematics 
of this system. The goal of this survey is to determine the membership and
masses of the three clusters, and whether they form a bound system.

Previous studies of Hercules galaxies can be grouped into 
three categories: studies of the region containing the supercluster, 
studies of the richest cluster, Abell~2151, and studies of the supercluster 
itself. Studies of the region include work by Freudling and 
collaborators (1988, 1991), who used the Tully-Fisher relation 
to study peculiar 
velocities, and Maccagni, Garilli, \& Tarenghi (1994), who studied the galaxy 
distribution with optical data. Giovanelli, Chincarini \& Haynes (1981) and
Dickey (1997) studied the supercluster in HI, concluding that there were
strong environmental effects on the mass of neutral hydrogen in the supercluster 
galaxies. Several groups (\cite{dre88}; \cite{bir95}; \cite{hua96}) have 
studied the structure of 
A2151 with the optical galaxy distribution and X-ray maps. All 
groups report the presence of substructure in the cluster.
The supercluster kinematics, especially involving A2147 and A2152,
have been less well-studied. The last major kinematical study was done by 
Tarenghi et al.\ (1979, 1980), using a total of 150 redshifts 
(124 in the supercluster) in a 28 square 
degree field. They concluded that it was difficult to separate the 
galaxies into three clusters unambiguously and that, contrary to what is 
seen in other rich clusters, the Hercules regions dominated by early-type 
galaxies showed little evidence for gravitational relaxation.

The three clusters that form the Hercules supercluster have both interesting 
differences and similarities. All three are classified as Bautz-Morgan
type~III (\cite{bau70}; \cite{lei77}) and Rood-Sastry type~F (\cite{str82}). 
All three are irregular and have spiral fractions of~$\sim$~50\%; 
A2147 is the most regular of the three and has the lowest spiral fraction 
(\cite{tar80}). A2147 and A2151 both 
have cooling flows (\cite{hen96}; \cite{hua96}). Henriksen (1992) reported 
that A2151 and A2152 have similar X-ray luminosities ($\sim 6.3 \times 
10^{42}h^{-2}$ erg~s$^{-1}$),\footnote{We use $H_0=100h$ km~s$^{-1}$~Mpc$^{-1}$.}
while A2147 has a much larger luminosity ($1.8 \times 10^{44}h^{-2}$ erg~s$^{-1}$).
Ebeling et al.\ (1996), using ROSAT data, explain some of this discrepancy
by showing that most of the X-ray flux from A2147 is from an AGN in the cluster 
and not its ICM. They give the X-ray luminosities of A2147 and A2151 as 
$7.0 \times 10^{43}h^{-2}$~erg~s$^{-1}$ 
and $2.4 \times 10^{43}h^{-2}$~erg~s$^{-1}$, respectively. 
The clusters all have unusual individual characteristics: A2151 contains
well-known substructure (\cite{bir95}), A2147 contains ``an unusual diffuse
radio source'' (\cite{bur94}), and A2152 has by far the highest velocity 
dispersion (${\sigma}_v = 1346$~km~s$^{-1}$) in the study by Zabludoff 
et al.\ (1993b) of dense cluster cores.

\section{Observations and Data Reduction}

Our `Hercules supercluster' region of interest, comprising 22.5 square degrees
($14\arcdeg 30\arcmin < \delta < 19\arcdeg 30\arcmin$ 
and $15^{\rm h} 54^{\rm m} <\alpha < 16^{\rm h} 12^{\rm m}$, B(1950)), 
now has a total of 294 velocities available from 
the literature. Of these, approximately 262 are possible cluster members 
($8500 < v < 14500$ km~s$^{-1}$). To augment this data, we constructed a 
catalog of galaxies in the region by combining and comparing two catalogs: one 
from the Minnesota Automated Plate Scanner (APS) scans of the Palomar Sky Survey 
(\cite{pen93}) and one we generated using FOCAS (\cite{val82}) 
and the Digitized Sky Survey (\cite{las91}). The catalog was 
magnitude-calibrated using $B$ and $R$ CCD images of portions of the region 
obtained at the 1.2m telescope of the Whipple Observatory on Mt. Hopkins (see 
Appendix for more details).

We measured new redshifts from the magnitude-ordered version of our Hercules
catalog, using the 1.5m telescope at FLWO with the
FAST Cassegrain spectrograph (\cite{fab97}), a 300~l~mm$^{-1}$ grating, and a CCD 
detector. Integration times were typically 10 to 20 minutes, and data reduction
was carried out with standard cross-correlation techniques (\cite{kur92}).
We also obtained some new redshifts from a separate study of the Hercules
$K$-band luminosity function (\cite{hu97b}). All the data for galaxies in the 
region is in Table~\ref{all-dat}, where the columns are (1) name, (2) RA (J2000),
(3) declination (J2000), (4) heliocentric velocity in km~s$^{-1}$, (5) velocity
error, (6) $R$~magnitude (from the CCD calibration described in the
appendix), (7) morphological type, and (8) velocity source.\footnote{
Table~\ref{all-dat} is available from the authors in electronic form.}
A velocity error of 100 km~s$^{-1}$ was assumed for velocities without 
published errors. Our redshift sample is complete to an $R$ magnitude of 15.1; 
our Hercules catalog of 293 galaxies without previously measured redshifts has 
a magnitude limit of $R=15.9$. 

\begin{figure}
\plotone{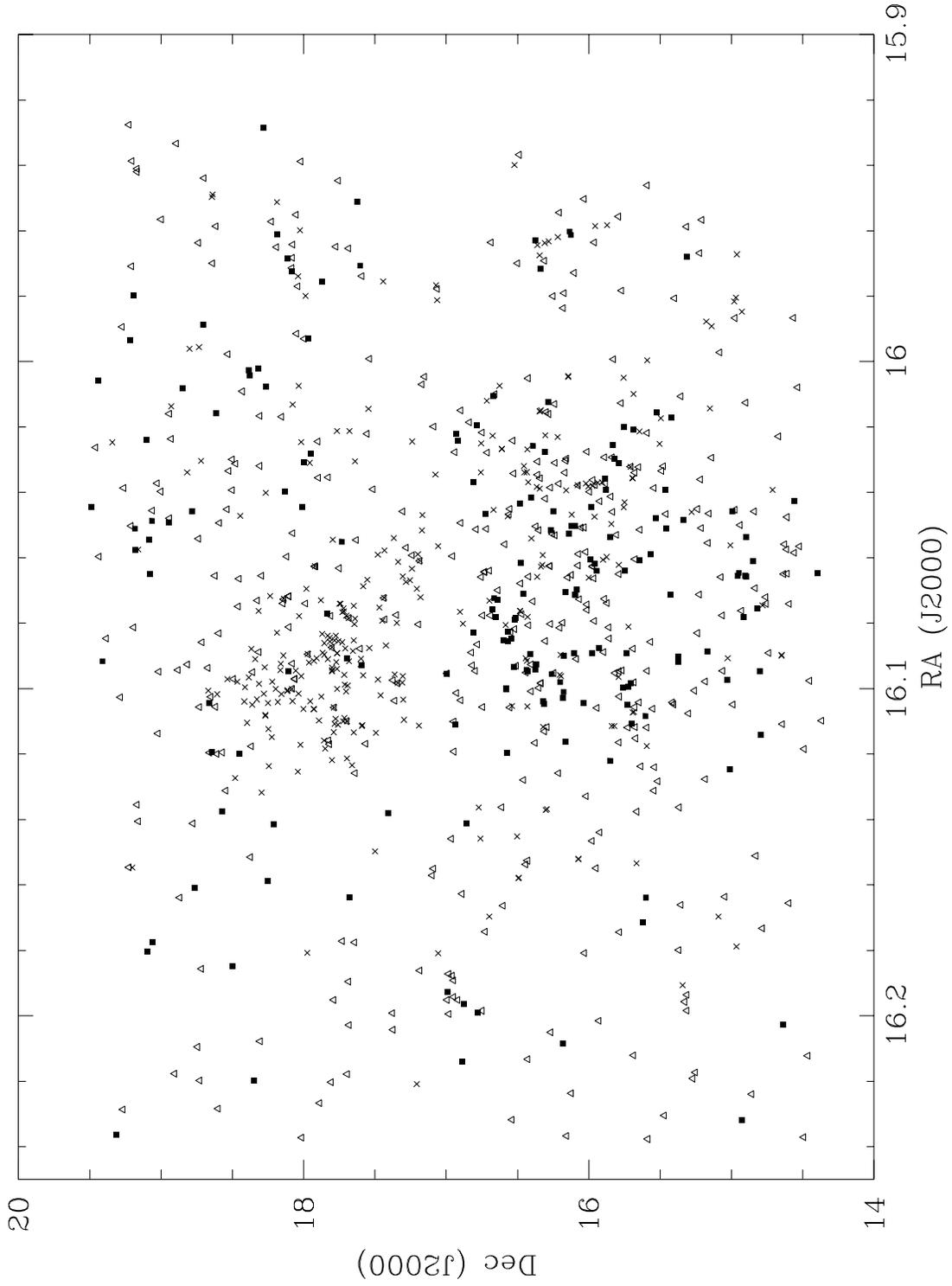}
\caption{Hercules supercluster field. Crosses: galaxies
 with velocities available from the literature, squares: galaxies with
 newly measured velocities, triangles: galaxies with $R<15.9$ and
 unmeasured velocities.\label{vel-pos}}
\end{figure}

\begin{figure}
\plotone{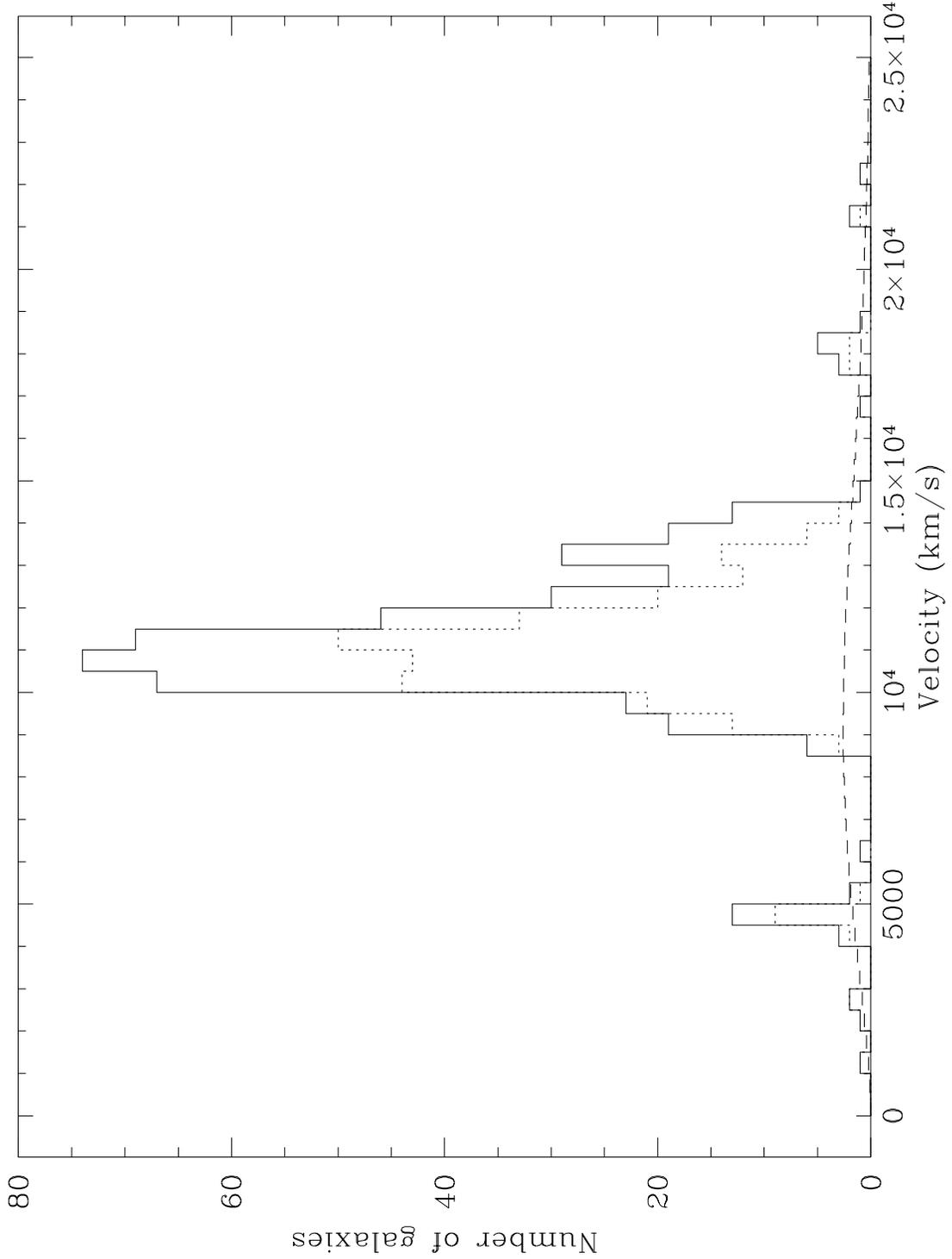}
\caption{Dotted line: Hercules supercluster velocities from the literature. Solid
  line: literature velocities and new velocities from this work.
  Dashed line: selection function (expected number of galaxies given 
  field luminosity function and survey magnitude limit).\label{new-old-hist}}
\end{figure}

\begin{figure}
\plotone{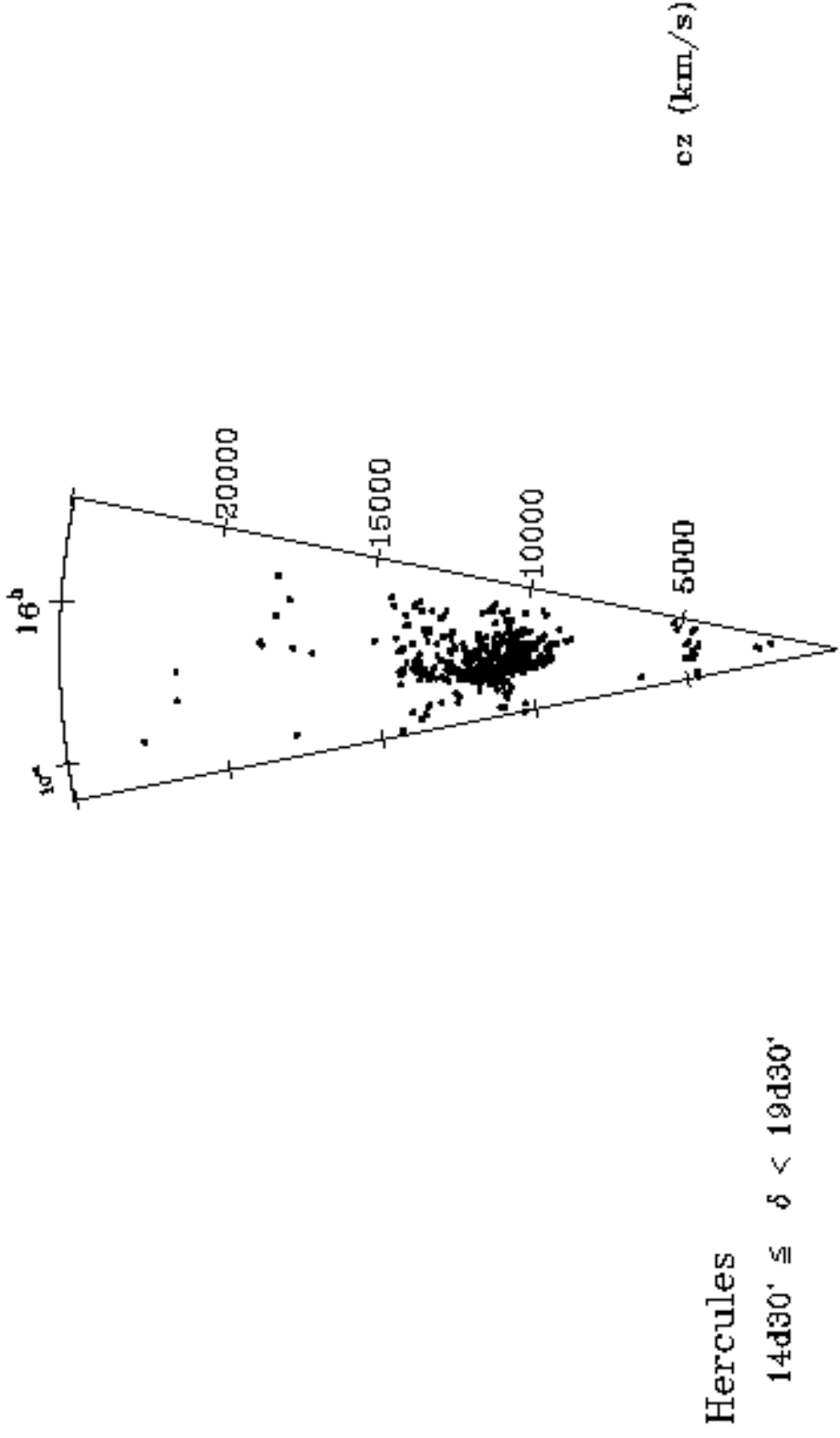}
\caption{RA cone diagram for galaxies in the Hercules Supercluster field\label{ra-cone-herc}}
\end{figure}

\begin{figure}
\plotone{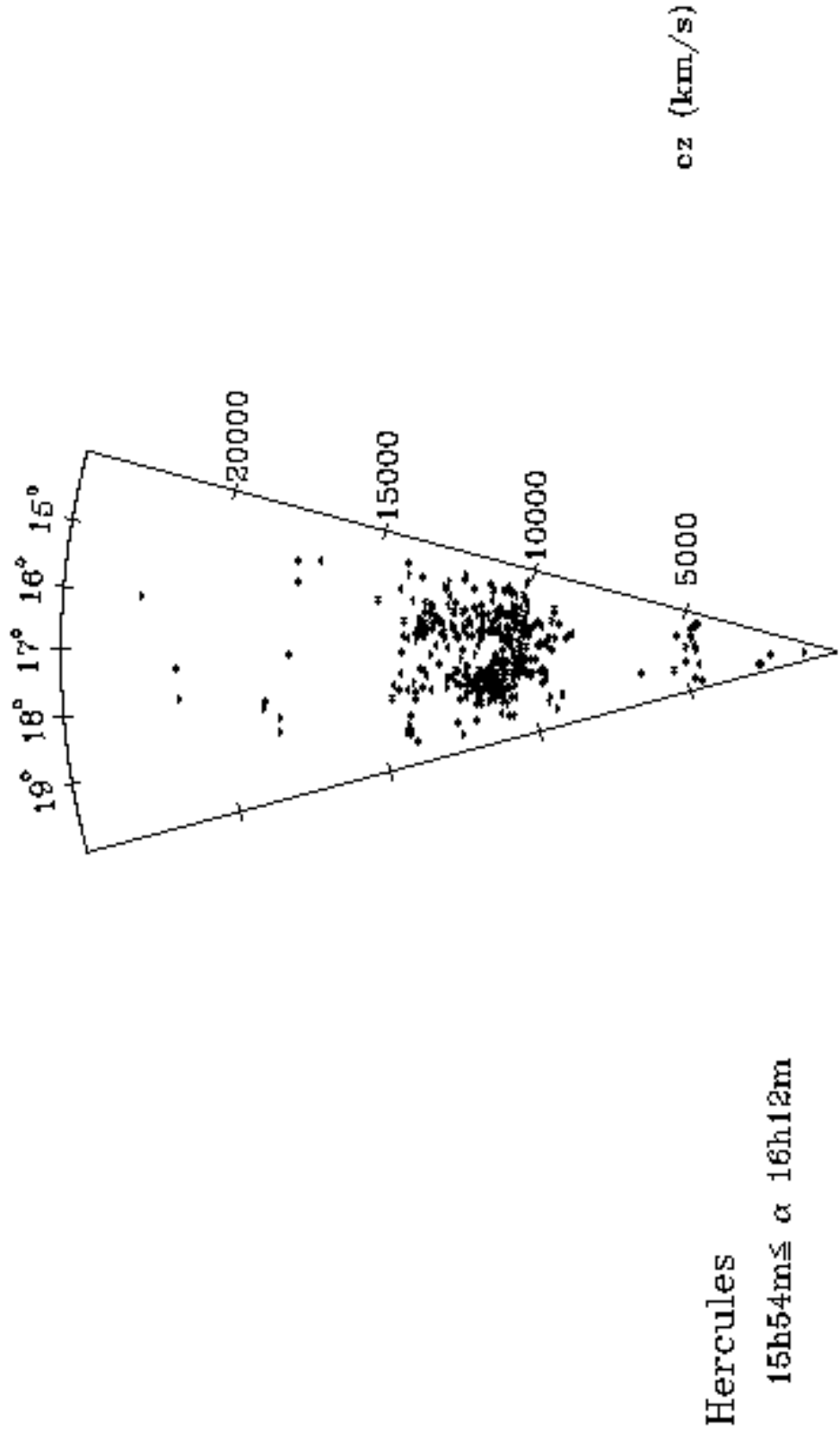}
\caption{Declination cone diagram for galaxies in the Hercules Supercluster field\label{dec-cone-herc}}
\end{figure}

Of the total of 175 new redshifts, 152 belong to possible cluster members; 
thus we can do a kinematical study of the supercluster with more than three times 
as many redshifts as Tarenghi et al.\ (1980). 
Figure~\ref{vel-pos} shows the positions of the galaxies with new and literature 
velocities; most of the new velocities
are in A2147, A2152, or the ``dispersed'' supercluster, since A2151 has been 
extensively observed. Figure~\ref{new-old-hist} shows the 
distribution in velocity of the literature and new velocities. 
The galaxies with new velocities in these bins
have a similar distribution on the sky (within the limits of
small-number statistics) to the galaxies with literature velocities
in the same bins. The cone diagrams in Figures~\ref{ra-cone-herc}
and \ref{dec-cone-herc} show all velocities in our field; from this it can be seen 
that the Hercules supercluster suffers from relatively little
foreground or background contamination. 

\section{Cluster Separation}

Tarenghi et al.\ (1980) noted the difficulty in separating the 
Hercules supercluster into the three Abell clusters. The presence of 
X-ray gas in all three clusters (\cite{hen92}) suggests that they are 
separate dynamical entities; we wanted to confirm this using our position 
and velocity information before trying to separate the clusters. 
The most obvious test is to find out whether 
the velocity distribution is composed of a single Gaussian. 
We did this using a Lilliefors test, a variant of the K-S test in which 
the parameters of the Gaussian to be compared to the velocity distribution
are derived from the distribution itself.
We also computed several indicators of substructure: the skewness and kurtosis
of the velocity distribution, and several statistics which measure substructure
by computing the values of a quantity for each galaxy and its nearest neighbors.
The $\Delta$ statistic of Dressler \& Shectman (1988b) measures the deviation 
of the average velocity and velocity dispersion computed for each galaxy and its 
nearest neighbors in projection from the global average velocity and global 
velocity dispersion. The $\alpha$ statistic of West \& Bothun (1990) measures the
deviation of the position centroids computed for each galaxy and its 
nearest neighbors in velocity from the global centroids.
The $\epsilon$ statistic of Bird (1994) combines position and velocity 
information by computing the projected mass estimator (see 
Section~\ref{sec-clpar}) 
for each galaxy and its nearest neighbors in projection.

We performed a Lilliefors test on the velocities of all objects in the supercluster 
using the ROSTAT statistics package of Beers et al.\ (1990); a Gaussian
was rejected at the 99\% level. 
We computed the substructure statistics following the procedure 
outlined by Bird (1994); the number of nearest neighbors used was 
equal to the square root of the number
of galaxies. Unlike Bird, we used the standard mean and dispersion
estimators, instead of biweight estimates.
We computed the significance of the statistics by comparing the value of 
the statistic computed for the cluster to values computed for Monte Carlo 
realizations 
of the cluster generated by scrambling the velocities of the galaxies. The
significance is the fraction of Monte Carlo realizations that have a value of
the statistic less than that computed for the cluster.
The significance of the skewness and kurtosis were evaluated by noting that 
their values are equal to the probability that a Gaussian distribution would have 
the same skewness or kurtosis as the observed distribution.
The substructure statistics for the supercluster are shown in 
Table~\ref{substruct}. All statistics except the $\epsilon$ test
indicated the presence of substructure, significant at the ($>90$ \%) level. 
From the Lilliefors and substructure tests we easily conclude 
that the supercluster
is not a single dynamical entity, and that it is reasonable to attempt to
assign galaxies to clusters.

\subsection{Techniques}
Several approaches can be used to assign cluster memberships and probabilities.
Two well-known cluster-finding techniques are ``friends-of-friends'' (\cite{huc82}),
which is a percolation algorithm, and the minimal spanning tree (\cite{bar85}). 
The ``friends-of-friends'' algorithm finds companions
of a galaxy and then the companions of the companions. All galaxies connected
to the initial galaxy in this way are part of one cluster. The separations in
position and velocity ($V_0, D_0$) at which
two galaxies are considered to be companions change with the magnitude limit
of the survey and the distance of the galaxies. The minimal spanning tree (MST)
of a dataset is the shortest graph which connects all objects in the set with no
circular paths; there are several simple algorithms for constructing such a
structure. With the MST in hand, clusters can be constructed by ``separating''
the tree -- that is, cutting ``branches'' longer than a certain length. Objects
still connected after separation are part of the same cluster.

Neither of the above methods produces membership probabilities, although 
these could be assigned for friends-of-friends using density parameters. 
Two methods were explored that do assign probabilities: the KMM (
``Kaye's Mixture Model'') algorithm 
(\cite{ash94}) and fuzzy clustering (\cite{jai88}; \cite{kau90}). 
The KMM algorithm 
defines the probability of an object's membership in a cluster as the 
Gaussian distance from the object to the cluster center:
\begin{equation}
f_i= \exp\left(-\left(\frac{\alpha-{\alpha}_c}{2{\sigma}_{\alpha}}\right)^2 -
\left(\frac{\delta-{\delta}_c}{2{\sigma}_{\delta}}\right)^2 
- \left(\frac{v-v_c}{2{\sigma}_v}\right)^2\right)
\end{equation}
properly normalized by the sum of its membership probabilities in all clusters.
It fits a user-specified number of Gaussian clusters to the data, 
maximizing a likelihood function based on the membership probabilities. 
The user must supply an initial guess for the locations of the clusters, but the
final result does not depend sensitively on this guess (see below).
Final membership probabilities are calculated after the KMM algorithm has converged. 
Fuzzy clustering also requires that the user specify the number of clusters, $m$, but
calculates membership probabilities directly, without first dividing the objects
into clusters. The goal is to minimize the objective function
\begin{equation}
C={\sum}_{k=1}^m \frac{{\sum}_{i,j=1}^n {u_{ik}}^2 {u_{jk}}^2 d_{ij}}
{2{\sum}_{j=1}^n {u_{jk}}^2} 
\end{equation}
where $u_{ik}$ is the membership probability of object $i$ in cluster $k$ and
$d_{ij}$ is the distance between objects $i$ and $j$ (a sum of
projected distance on the sky and line-of-sight velocity difference weighted
by a factor $w$), i.\ e.\ 
\begin{equation}
d_{ij}=\left(\left({\theta}_{ij} \frac{v_i+v_j}{2}\frac{1}{H_0}\right)^2+ 
\left(w\frac{v_i-v_j}{H_0}\right)^2\right)^{1/2} 
\end{equation}
To minimize the objective function, we used the algorithm given 
in Kaufmann \& Rousseeuw (1990), which iteratively finds the local minimum
using the method of Lagrange multipliers.

While there have been some tests of these individual methods against simulations
(e.\ g.\ \cite{bar85}; \cite{ash94}) there is little information available
on their comparative performance in the context of separating nearby clusters.
We therefore tested these methods on simulated clusters made to resemble 
our actual 
data. Clusters were simulated by picking galaxies' velocities at random 
from a Gaussian distribution and their positions at random from a 
truncated King model for the surface density. Velocity dispersions ranging 
from 700 to 1000~km~s$^{-1}$ and core radii from  0.45-0.55~Mpc, which
are typical for the Hercules clusters, were used. 
A magnitude-limited background was also included. We simulated fields of 
three clusters, varying the distance between clusters in position and 
velocity space using a 25-model grid with 5 different values each of position
and velocity separation.
Since all of the methods require user input parameters we tuned the 
performance of each algorithm by using the best result from a range of 
parameters.

In order to quantify the accuracy of the various separation methods,
we defined a ``separation statistic'' $S$, to be calculated for
each of our simulated cluster sets. The algorithm for calculating
$S$ is as follows:
\begin{enumerate}
\item Calculate the centers of each cluster found by the cluster-finding
method.
\item For each method, determine which `found' cluster corresponds to 
each original input cluster (the one closest in position on the sky).
\item Calculate the ``correctness'' $c_i$ for each galaxy in each method.
If the cluster the galaxy was assigned to by the method corresponds to its
original input cluster, the ``correctness'' is 1; otherwise it is 0.
For KMM and fuzzy clustering, each galaxy is assigned to the cluster for
which it has the greatest membership probability.
\item Calculate $S$ as the sum of $c_i$, normalized by the total 
number of galaxies:
\end{enumerate}
\begin{equation}
S= \frac{1}{N}{\sum}_i^N c_i
\end{equation}
The $S$-statistic can thus be considered a count of the ``correct'' answers.
The background galaxies were used in the cluster-finding methods but were not
included in the calculation of the statistic.

\begin{figure}
\plotone{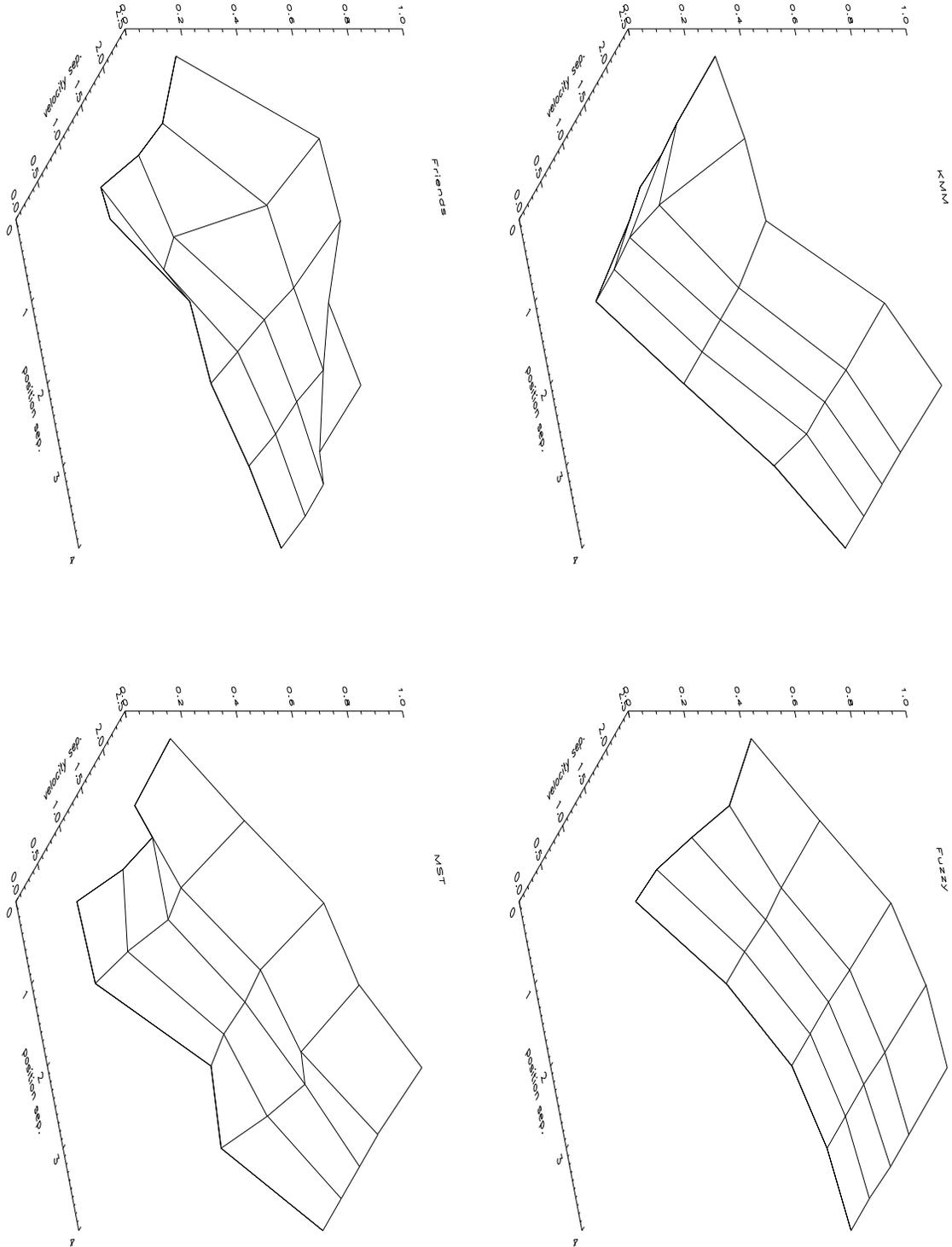}
\caption{Separation statistic as a function of position and velocity separation
for the four cluster-finding methods. \label{separate}}
\end{figure}

Figure~\ref{separate} shows the values of the separation statistic for
all four methods as a function of average position and velocity
separations. The low values for all methods at low separations
reflect the fact that no method could effectively separate the clusters
in the region ($\bar{\Delta \theta} / {\theta}_{\rm cl} < 1$, 
$\bar{\Delta v}/{\sigma}_v<1.0$). 
Overall, KMM and fuzzy clustering are the better performers.
As expected, the performance
of all the methods generally improves as the average separation increases; the
improvement is larger for the angular separations. This is probably
because the King model for the spatial positions of the galaxies
is more centrally concentrated than the Gaussian used for their velocities.
Friends-of-friends  performs poorly as separation increases 
due to fragmentation: with input parameters such that the clusters were
separated from each other, they were also separated into smaller pieces.
This suggests that friends-of-friends is probably better suited
to finding well-separated and compact clusters in data (as was its original
purpose) than separating nearby clusters.

To compare fuzzy clustering and KMM we calculated another statistic.
For the purposes of this study we were interested in determining
which method of assigning galaxies to clusters produced the best estimate 
of the properties of the clusters, not just in determining which galaxy
belongs to which cluster. To quantify this property of the separation 
methods, we calculated the six parameters $x_i$
(mean RA, dec, velocity, velocity dispersion, virial mass,
and projected mass) of each 
cluster using the membership probabilities as weights, as described in
Section~\ref{sec-clpar}. Then we compared these to the input parameters
used to generate the clusters and calculated the RMS residuals, e.\ g.\
\begin{equation}
R = \sqrt{\frac{1}{6N_{\rm cl}-1}\sum_{\rm cl}\sum_{i=1}^6 
 \left(\frac{{x_i}_{\rm cl}-{x_i}_{\rm inp}}{{x_i}_{\rm inp}}\right)^2}
\end{equation}
The first sum is over the 3 ``clusters'' found by the algorithms;
the second is over the 6 parameters computed for each. A lower
value of the statistic represents better performance. (We note that this
statistic should not be used to compare the results of clustering methods
on clusters generated with different input parameters. 
For example, consider two sets of input clusters, where set $A$ is more
widely separated in velocity than set $B$. 
Even if the clustering results are equally accurate for both sets,
the incorrectly assigned galaxies in the results for set $A$ will be more likely 
to have velocities further from the mean and hence bias the
calculated velocity dispersions upward.)

From the results of Tarenghi et al. (1980) and Zabludoff et al.\
(1993b), we estimated the average
separations of the three cluster pairs to be approximately 1.4\arcdeg \
and 350-550~km~s$^{-1}$. For the artificial clusters closest to these
parameters, KMM had the better $R$-statistic in each case. This was
because fuzzy clustering's results are ``too fuzzy'': although the
galaxies are usually assigned to the correct cluster, they still
have significant membership probabilities (up to $\sim 0.3$) in the clusters 
to which they are not assigned. (While this may be appropriate for
a few galaxies for which the cluster assignment really is uncertain, 
it is physically unreasonable for most galaxies to ``belong'' to more
than one cluster.) The overly fuzzy assignments result in the mean velocities
of the three clusters being biased toward the mean of all the
velocities, and the velocity dispersions being biased upward.

These two statistics show that the performance of various clustering methods
depends heavily on the separation of the input clusters. The method
to be used should depend on the problem at hand. Based on the
results of our statistics, and for the Hercules Supercluster,
the KMM method is the best for our purposes.

We tested the robustness of the KMM cluster-assignment algorithm using a
jackknife procedure on our supercluster data. The KMM input data 
(galaxy positions and velocities) were randomly ordered, and the KMM
algorithm was run on subsets of the original data. We fit three clusters 
to the data, corresponding to the three Abell clusters. 
The subset sizes were 
linearly increased from 100 galaxies to the full dataset
of galaxies used in later analysis.
The KMM input parameters (such as the initial cluster positions) 
were kept constant. The jackknife results showed only 
a small amount of scatter in the central positions and velocities
of the clusters. The largest scatter was,
as might be expected, for A2152, the cluster with the fewest galaxies and
the largest spatial and velocity dispersion. Even so, the dispersion of
estimated central velocities was only 112~km~s$^{-1}$, and the largest 
deviation of estimated central position
about 12$\arcmin$. We also tested the KMM procedure by
using a range of initial cluster positions and velocities, similar to the 
procedure used by Colless \& Dunn (1996). We found results similar to theirs -- 
namely, that KMM converged to similar results with either good initial cluster
velocity estimates (within 2000~km~s$^{-1}$) or initial positions (within
30\arcmin). The algorithm failed to converge to these clusters only when
both position and velocity information were omitted. From these
tests we concluded that the KMM results should be robust.

\subsection{Results}\label{cl-assign-res}

\begin{figure}
\plotone{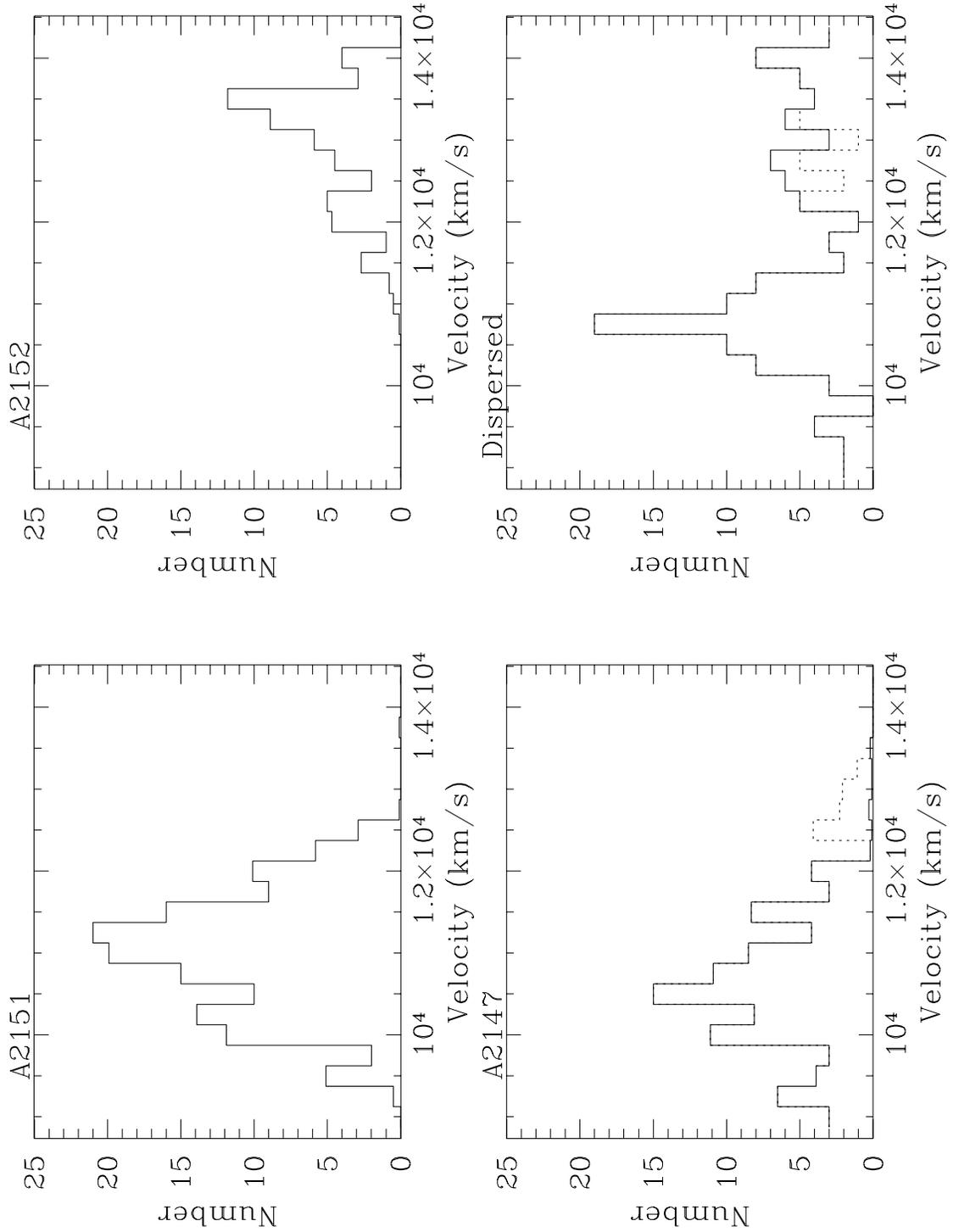}
\caption{Velocity histograms for A2151, A2152, A2147 and the dispersed component. 
The dotted line shows the effect of removing the background group
from A2147. \label{vel-hist}}
\end{figure}

The KMM algorithm is sensitive to outliers (\cite{bir95}), so we restricted our
final analysis to galaxies within 0.85\arcdeg \ (1.6$h^{-1}$ Mpc) of the 
projected cluster 
centers. To find these centers we assigned each galaxy to the nearest of the
Tarenghi et al. (1980) cluster centers-of-mass (A2151: 16$^{\rm h}$ 5$^{\rm m}$ 
26$^{\rm s}$, 
17\arcdeg 47\arcmin 50\arcsec, A2152: 16$^{\rm h}$ 5$^{\rm m}$ 6$^{\rm s}$, 
16\arcdeg 19\arcmin 41\arcsec, A2147: 16$^{\rm h}$ 1$^{\rm m}$ 59$^{\rm s}$, 
16\arcdeg 2\arcmin 55\arcsec (J2000)) and then recalculated the centers 
using only galaxies
within 0.85\arcdeg\  of the center. This process was iterated until there
was no further change in the center location; this typically only required
three or four iterations. The resulting dataset had a total of 301 galaxies 
and did not include outlying groups which might bias the KMM solution.
The remaining 113 galaxies in the supercluster velocity range 
(the ``dispersed component'' in the terminology of Tarenghi et al.\ (1980)) 
were analyzed kinematically but not fit to clusters. 
Resulting velocity histograms for the solutions are in Figure~\ref{vel-hist}. 

After inspecting the KMM results, we made one change to the cluster
assignments.
KMM assigned nine galaxies with velocities greater than 12000~km~s${^-1}$
to A2147. At first glance this seems reasonable: since these galaxies are 
on the west side of A2147, they are unlikely to be part of A2152,
a degree away on the sky. We decided after inspecting the velocity
histogram (Figure~\ref{vel-hist}), however, that these galaxies were more 
likely part of the dispersed supercluster or a separate background group
than of A2147, so we reassigned them. 
This reassignment also makes the velocity dispersion of A2147 more
compatible with its measured X-ray temperature (see Section~\ref{sec-clpar}).
After the reassignment there are 293 cluster galaxies and 122
members of the dispersed supercluster.

\begin{figure}
\plotone{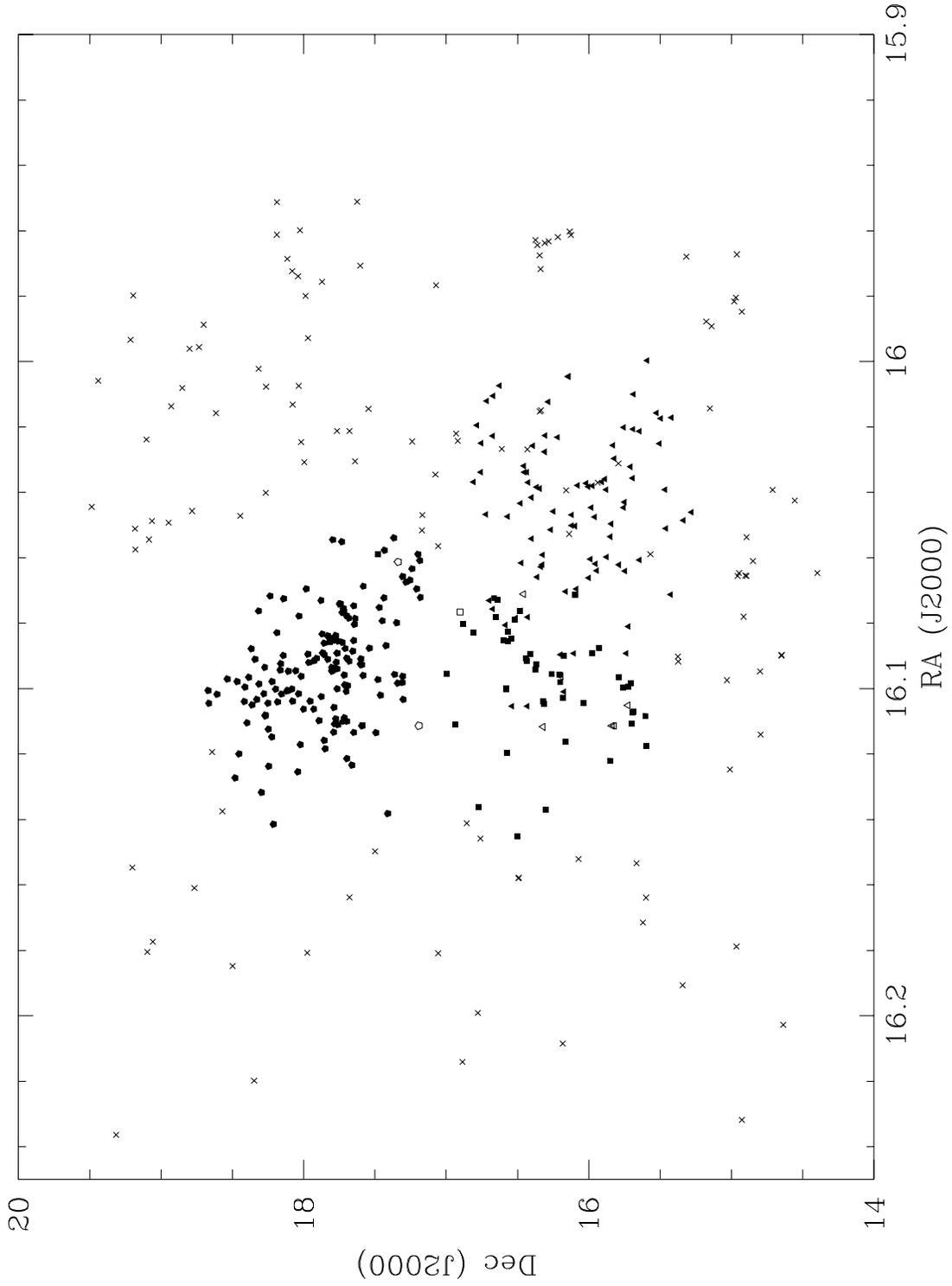}
\caption{Cluster assignments. A2151: filled pentagons, A2147: filled triangles,
A2152: filled squares, A2147 and A2152: open triangles, A2151 and A2147: open squares, 
A2151 and A2152: open pentagons, dispersed component: crosses.
\label{cl-assign}}
\end{figure}

Our computed
parameters for A2151 (see Table~\ref{cl-param}) were similar to other
reported values. This is not surprising, as it is reasonably well-separated 
from the other two clusters, has the most available velocities, and
had relatively few new velocities added.
However, our solutions for the separation of A2152 and A2147
were different from previously reported results. A2152 had a significantly higher 
mean velocity than previously reported (12942$\pm$97~km~s$^{-1}$, 
error computed using formulas in Section~\ref{sec-clpar}), 
and both A2147 and A2152 had lower velocity dispersions 
(821 and 715~km~s$^{-1}$), as compared to Zabludoff et al.'s 
1081~km~s$^{-1}$ (A2147) and 1346~km~s$^{-1}$ (A2152). 
About 15 galaxies 
had significant membership probabilities ($>20$\%) for more than one cluster; 
these galaxies were fractionally assigned
to the appropriate cluster. Thus, in the velocity histograms, the number 
of galaxies in each bin is not necessarily an integer.
Figure~\ref{cl-assign} shows the cluster assignments of all 292 galaxies for the 
KMM solution and the locations of the 122 members of the ``dispersed 
supercluster''.

\section{Cluster Dynamics} 

\subsection{Cluster parameters}\label{sec-clpar}

We computed the usual cluster parameters for all three of our cluster assignment 
solutions. In computing the mean velocities and dispersions given in 
Table~\ref{cl-param}, fractionally assigned
galaxies were included in the calculation for a cluster weighted by 
their membership probabilities, $f_i$, e.g.
\begin{equation}
\bar{v} =\frac {\sum_{i=1}^n f_i v_i}{\sum_{i=1}^n f_i} 
\end{equation}
\begin{equation}
{\sigma}_v=(\frac{\sum_{i=1}^n f_i (v_i-\bar{v})^2}{\sum_{i=1}^n f_i-1})^{1/2} 
\end{equation}
$1\sigma$ confidence levels for these parameters were computed 
using the formulas of Danese,
De Zotti, and di Tullio (1980), modified to take the membership probabilities into
account: 

\begin{equation}
(\Delta \bar{v})^2 = c^2 \frac{k^2({{\sigma}_v}'/c)^2 + \bar{{\delta}^2}/c^2} {n'}
\end{equation}

\begin{equation}
(\Delta {\sigma}_{\pm})^2 = \left(\left(\frac{n'-1}{{\chi}_{\pm}(k)}\right)^{1/2} -1\right)^2 ({{\sigma}_v}')^2 
  + \frac{\bar{{\delta}^2} \left(1+\frac{2{{\sigma}'_v}^2 
\bar{{\delta}^2}}{(1+\bar{v}/c)^2}\right)}{n' (1+\bar{v}/c)^2}
\end{equation}

where
\begin{equation}
n'={\sum}_i f_i \quad \quad \bar{v}=\frac{{\sum}_i f_i v_i}{n'} \quad \quad 
  \bar{{\delta}^2}=\frac{{\sum}_i f_i {\delta}_i^2}{n'}
\end{equation}

and 

\begin{equation}
({{\sigma}_v}')^2=\frac{1}{c} \left(\frac{{\sum}_i \frac{f_i (v_i-\bar{v})^2}
{1+\bar{v}/c}}{n'} 
  - \frac{\bar{{\delta}^2}}{(1+\bar{v}/c)^2}\right)
\end{equation}
$\chi_{\pm}(k)$ is a numerical factor which depends on the confidence level 
$k$ and the number of measurements $n$. 
Errors in the weights were not taken into account in our modifications of these 
formulae.

We obtained few new velocities in the region of A2151, and, as expected, our
mean velocity and dispersion results were compatible, within the error, to 
previously published results. However, our results for A2152 and A2147 were 
quite different from previous results: we found lower velocity dispersions, 
of 715 and 821~km~s$^{-1}$, 
and a greater mean velocity separation between the two clusters, mostly 
due to an increase in the mean velocity of A2152. We redid the KMM analysis 
without our
new velocities and found, using a Student's $t$-test, that the increase in 
mean velocity between the solution derived with the new velocities and the one
without was statistically significant at the 99.3\% level.

We estimated the masses of the individual clusters using two standard methods:
the virial theorem and the projected mass estimator of 
Heisler, Tremaine, \& Bahcall~(1982). The contribution of each galaxy to the mass 
of a cluster was weighted by its membership probability $f_i$ in that cluster. 
For example, we used the following expression for the virial mass of a cluster:
\begin{equation}
M_{\rm VT}=\frac {3 \pi ({\sum}_i f_i) {\sum}_i f_i  {v_i}^2}
{2G {\sum}_{i,j} f_i f_j/r_{ij}}
\end{equation}
Similar modifications were made for the 
projected mass. The results are in Table~\ref{cl-param}; the errors in the
masses are 1$\sigma$ confidence levels.
Previously determined masses for the Hercules clusters are in
Table~\ref{other-mass}; our masses for A2151 are in
general agreement with those of Bird, Davis \& Beers (1995) and 
Tarenghi et al.\ (1980). Our virial masses for A2147 and A2152 are smaller 
than those of Tarenghi et al.\ (1980)
by up to a factor of three, because we find lower velocity dispersions. 
Our projected masses are larger than the virial masses and therefore closer to 
the virial masses of Tarenghi et al. This is not due to our larger cluster membership 
cutoff; when we used their value of 0.8\arcdeg\ and re-calculated the masses,
the projected masses decreased by less than 15\% and the ratio of virial to 
projected mass changed by less than 5\%.

We also compared our computed velocity dispersions with 
X-ray data on the clusters. The compilation of David et al.\ (1993)
contains X-ray temperatures for A2151 and A2147; however, we were unable to 
find a published X-ray temperature for A2152. We use the $T_X-{\sigma}$
relation of Girardi et al.\ (1996) ($\sigma=10^{2.53\pm0.04}{T_X}^{0.61\pm0.05}$;  
2$\sigma$ error bars) to compute X-ray predicted velocity dispersions.
For A2151 ($kT_X=3.8$~keV), the prediction is $765\pm131$~km~s$^{-1}$,
compatible with our measured value of 705$^{+46}_{-39}$~km~s$^{-1}$.
For A2147 ($kT_X=4.4$~keV), the prediction for
A2147 is $836\pm151$~km~s$^{-1}$, compatible 
with our measured value of 821$^{+68}_{-55}$~km~s$^{-1}$
(after the background is removed; without the background removed
our velocity dispersion is 992$^{+78}_{-63}$~km~s$^{-1}$). 
We conclude that our velocity dispersions are compatible with the 
available X-ray temperatures.
The X-ray-derived mass for A2147 from Henriksen \& White (1996),
$4.9^{+2.6}_{-1.0}\times10^{15}M_{\sun}$, is much larger than both
our mass and that of Tarenghi et al.\ (1980).
Henriksen \& White did not correct for the presence of the AGN
detected by Ebeling et al.\ (1996); contamination from this
source may have resulted in their overly large mass.

Because our sample goes to a fainter magnitude
limit, we have a total of 122 ``dispersed'' cluster members while 
Tarenghi et al.\ (1980) have 35. The velocity histogram for
the dispersed component is in Figure~\ref{vel-hist} with the cluster
histograms. Our results
for the kinematics of the dispersed population are fairly similar, however:
we find a mean velocity and dispersion of $\bar{v}=11639\pm$128~km~s$^{-1}$ and 
$\sigma=1407^{+100}_{-83}$~km~s$^{-1}$, 
while they find a mean and dispersion of $\bar{v}=11216\pm$238~km~s$^{-1}$ 
and $\sigma=1407$~km~s$^{-1}$. We confirm their result that 
the velocity dispersion of the clusters
is smaller than that of the dispersed component. The only `substructure'
in the dispersed component is the group of galaxies at 15$^{\rm h}$~59$^{\rm m}$, 
16\arcdeg~12\arcmin, also observed in the Burns et al.\ (1987) survey of poor groups, 
and the possible background group to A2147 (see Section~\ref{vel-str}).

We estimated the mass-to-light ratios of the clusters by adding 
up the luminosities of the cluster galaxies and making a faint-end
correction. The luminosities were calculated from the $R$ magnitudes, 
using $R_{\sun}= 4.52$ (\cite{lin96}), $K$-corrections of the form 
$K(z)=-2.5\log(1+z)$, and corrections for galactic absorption from 
Burstein \& Heiles (1984), with $A_R=0.625 A_B$.
We derived the faint-end correction factor from the luminosity
functions derived by Lugger (1989) for A2147 and A2151 (we assume 
the luminosity function of A2152 to be similar to that of A2147). 
Although these functions were determined without the use of redshifts, they are 
derived from data which are background-corrected and have a faint limit
several magnitudes fainter than ours. We corrected the Schechter 
function parameters given by Lugger ($\alpha=-1.13, M_*=-22.60$, for 
A2147 and A2152, and $\alpha=-1.09, M_*=-23.00$ for A2151) to $h=1$;
the faint-end correction is then 
\begin{equation}
F=\frac{\Gamma(2+\alpha)}{\Gamma(2+\alpha,L/L_*)}
\end{equation} 
The total luminosity was then estimated
with the relation $L_{\rm TOT} = F \sum_i^{N_{\rm cl}} L_i$
The resulting total luminosities and mass-to-light ratios are in 
Table~\ref{lum-param}. We can compare these mass-to-light ratios
to the closure mass-to-light ratio, which is
\begin{equation}
\left(\frac{M}{L}\right)_{R, {\rm closure}} = \frac{3{H_0}^2}{8\pi G j}
\end{equation}
where $j$ is the field luminosity density: $j=\int_0^{\infty}L {\phi}_F(L) dL$.
Lin et al.\ (1996) give $j_R\approx 1.75 \times 10^8 L_{\sun}$Mpc$^{-3}$ which
yields $(M/L)_{R, {\rm closure}}= 1578 (M/L)_{\sun}$.
A2151 has a mass-to-light ratio well below this value, while 
A2152's ratio is about one third and A2147's ratio about one half of the closure 
density. This suggests either that these two clusters are extremely massive,
or that the masses are contaminated by the presence of substructure or
supercluster interlopers, or that our photometry has a serious
zero-point error we failed to detect.
These results are much larger than those of Postman, Geller, \& Huchra (1988), 
who derived a mass-to-light ratio of 256 in $R$ for the clusters in the Corona 
Borealis Supercluster.

\subsection{Velocity Structure}\label{vel-str}

Previous work has shown that the presence of substructure can significantly
affect the virial masses of clusters, and that A2151 has significant substructure
on scales of less than 1$h^{-1}$ Mpc (\cite{bir95}).
We attempted to detect the presence of substructure in the clusters.
Using a Lilliefors test, 
we found that the velocity distributions of A2147 differed significantly  
from a Gaussian. We also ran the substructure tests used 
on the supercluster on each of the individual clusters. For the purposes of 
these tests we assigned each galaxy to the cluster in which its membership 
probability was largest, since it was not obvious how to generalize these
tests for fractional membership. The results indicated the presence of substructure 
in A2151, significant at the 99.9\% level, for both the $\alpha$ and  
$\Delta$ statistics. A2152 also had significant (99.9\%) values of the $\alpha$    
and $\Delta$ statistics; however we are hesitant to claim detection 
of substructure since this result is based on only 56 redshifts. A2147 had   
a significant result for $\alpha$ and marginally significant one for $\Delta$;   
we regard this as tantalizing but again not convincing.
We tried running the KMM algorithm on A2147 and A2152, using several random guesses
for initial group positions and velocities, but the algorithms did not converge to
dynamically distinct groups. This illustrates that the initial guesses are 
important, and may require additional information such as X-ray maps and 
brightest galaxy positions. Using our data, the KMM algorithm, and 
the subcluster parameters of Bird, Davis and Beers (1995) as input to KMM, 
we were able to reproduce 
their KMM results for the substructure in A2151. The possible 
or definite presence of substructure in the clusters means that the virial
masses may be unreliable.

Velocity segregation between galaxies of different morphological type can also
be important in clusters. We separated all 292 galaxies used in the cluster 
fitting into elliptical/S0 and spiral 
groups based on morphological classifications from the literature (Dressler \&
Shectman 1988; Maccagni, Garilli \& Tarenghi 1995; Huchra 1996), 
or from our CCD frames and the Digitized Sky Survey images
if published classifications were not available. We calculated the mean velocities
and velocity dispersions for the ellipticals and spirals (Table~\ref{esvel}),
and tested for statistically significant (at the 95\% level) differences between the
two groups
(using a t-test for the mean velocities, and an F-test for the velocity dispersions).
We found differences in both the mean and dispersion for A2151 only.  
Tarenghi et al.\ (1980) found the same differences in A2151 (although 
the found the spirals to have a higher, rather than lower, velocity dispersion), 
and a difference in the dispersion of A2147. We do find a difference between the
ellipticals and spirals in A2147 when we add in the high-velocity galaxies
removed earlier (see Section~\ref{cl-assign-res}) since 2/3 of these are
spirals. 

The velocity differences in A2151 are probably accountable for by 
substructure, although
the results of other groups on the kinematics of the morphological-type groups
in A2151 confuse, rather than clarify, the question of which type has the larger
velocity dispersion. Zabludoff \& Franx (1993), separating the galaxies into three 
groups: (elliptical, spiral, and S0), found a significant difference 
only between the mean velocities of the ellipticals and spirals.
Bird, Davis \& Beers (1993; BDB) compared the kinematics of the three morphological 
groups in two substructures. They found the that spirals
had greater mean velocities, and that S0s had a far larger velocity 
dispersion in the central substructure. Maccagni, Garilli \& Tarenghi (1994; MGT)
compared E/S0 and S groups (for different substructures than BDB), 
finding higher mean spiral velocities for all three substructures 
and similar velocity dispersions for the NNE and central substructures.
They find a much higher velocity dispersion for their SSW substructure.
We conclude that our finding that the spirals in A2151 have a greater mean velocity
than the other types agrees with all of these groups and fits well with
the idea that presence of substructure in this cluster. 
The larger velocity dispersion we find for the E/S0 galaxies may be 
due to the large dispersion of the S0s found by BDB.

We were intrigued by one results quoted by MGT:
in their SSW substructure they found the spiral and elliptical/S0 velocity 
dispersions to be 1133~km~s$^{-1}$ and 142~km~s$^{-1}$. This is a difference  
of a factor of eight! However, there are only 8 ellipticals and 16 spirals
in this substructure, and they suggest the spirals may be part of a 
field structure. The `SSW' structure of MGT actually includes
the entire southern region of A2151. We calculated the dynamical parameters of
a SW structure defined as $16^{\rm h}3^{\rm m}<\alpha<
16^{\rm h}4^{\rm s}48^{\rm s}, 17\arcdeg 18\arcmin<\delta<17\arcdeg 30\arcmin$.
This group contains seven ellipticals and seven spirals. We found similar
Results to those of MGT (see Table~\ref{esvel}), with the ellipticals having a lower
mean velocity and a much lower velocity dispersion than the spirals; however,
the t- and F-tests did not show these differences to be statistically
significant. We did find statistically significant differences 
between the means and dispersions of the SW and A2151 spirals, and between the 
velocity dispersions of the SW ellipticals and the A2151 ellipticals,
and the SW galaxies and A2151 galaxies unseparated by type.
It is difficult to be certain given the small-number statistics, but the
evidence appears to point toward the existence of a separate group of
galaxies in the SW of A2151. This point will be revisited in 
Section~\ref{cl-class}.

\section{Galaxy and cluster morphology} 
\subsection{Cluster appearance in different galaxy type}

\begin{figure}
\plotone{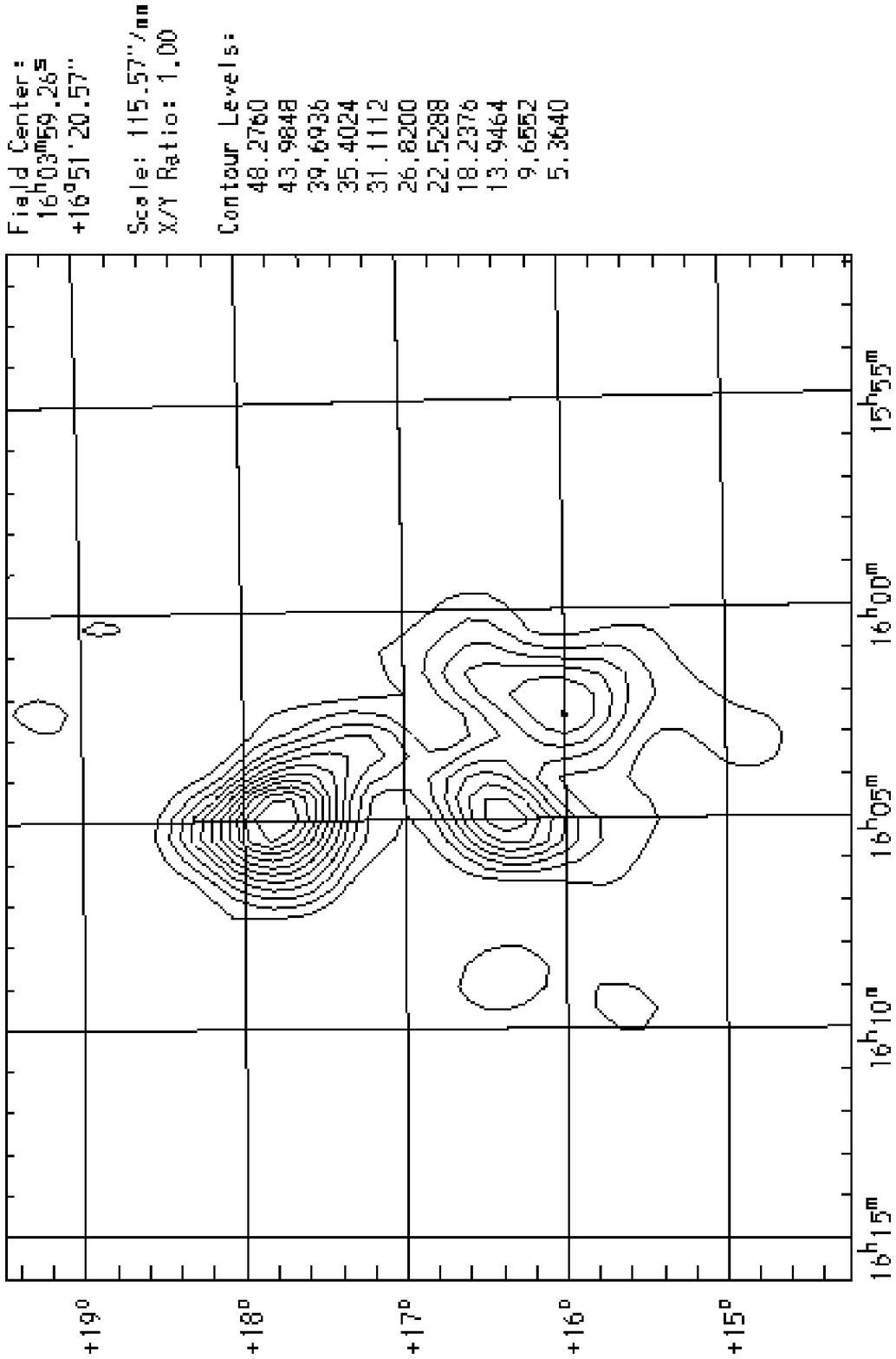}
\caption{Projected elliptical galaxy density; contours
are in units of galaxies per square degree, spaced linearly from 10 to
90 percent of the maximum value.\label{Econtour}}
\end{figure}

\begin{figure}
\plotone{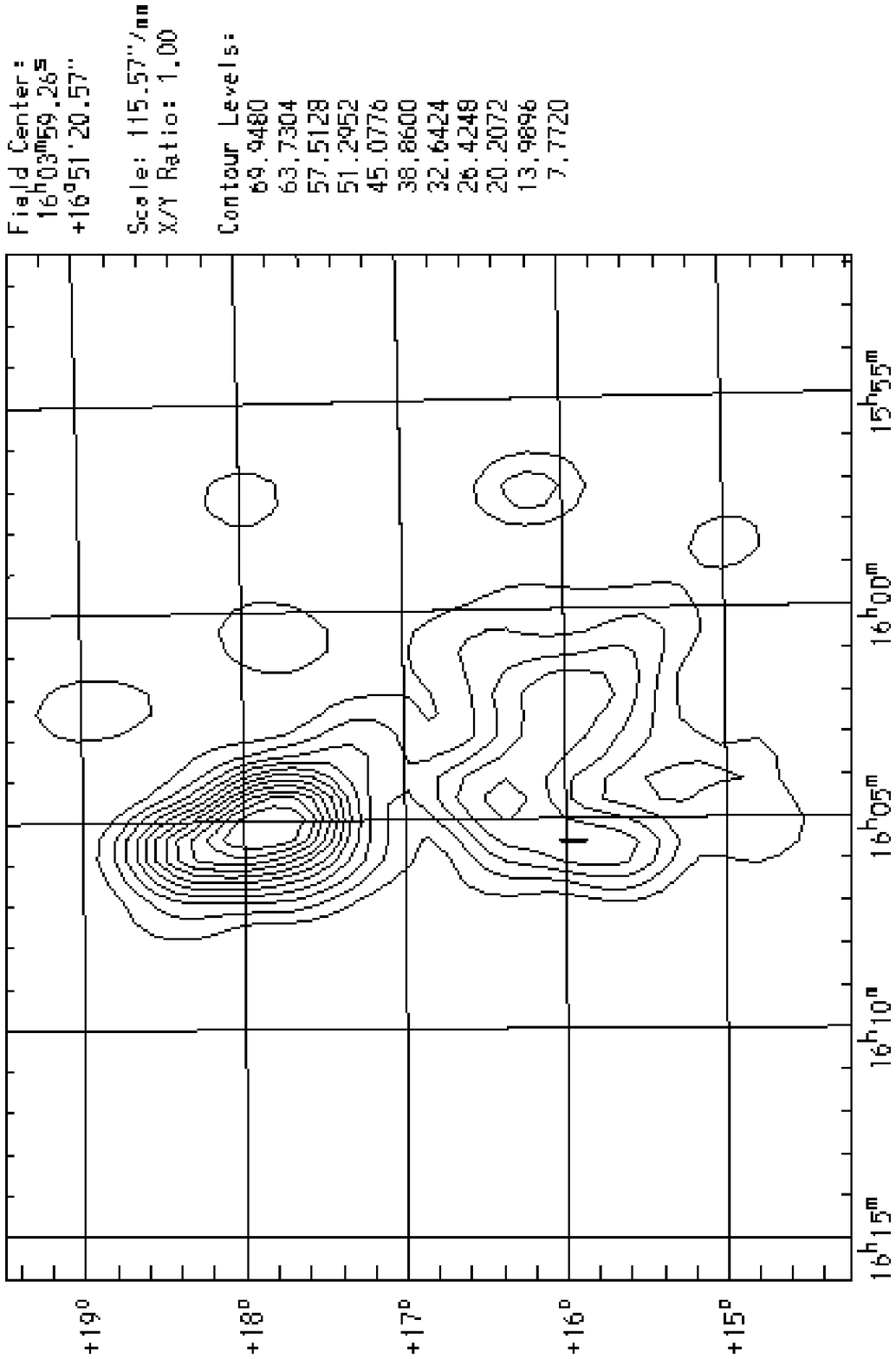}
\caption{Projected spiral galaxy density; contours
are in units of galaxies per square degree, spaced linearly from 10 to
90 percent of the maximum value.\label{Scontour}}
\end{figure}

With a large sample of galaxies, separated into clusters, and morphologically
classified, we can study the morphology-density relation in the three clusters.
The Hercules clusters have a much greater proportion of spirals than rich clusters like 
Coma and Virgo, so the near absence of spirals in the cluster center is not likely
to occur. Contour plots of the galaxy density in the supercluster 
(Figures~\ref{Econtour}~and~\ref{Scontour}) show something like the usual 
morphology-density relations for all three clusters: ellipticals are more 
concentrated in 
the center of the cluster. The centers and extent of the clusters in spirals and
ellipticals also appear to be similar.
Tarenghi et al.\ (1980) also found that the two types of galaxies 
had similar spatial distributions. Most of the density enhancements outside 
the three clusters consist mainly of spirals; the ``elliptical'' groups 
are smaller. This large spiral fraction is unsurprising, given the large spiral 
fraction in the supercluster overall.

\subsection{Cluster classification}\label{cl-class}
An interesting property of the Hercules clusters is their cluster
morphological type. The classification of the Hercules clusters as BM-III 
and RS-F would seem to be incompatible with the presence of cD galaxies,
which are supposed to form in rich, relaxed clusters. 
However, Zabludoff et al.\ (1993b) classify both A2147 and A2151 as cD 
clusters with the cDs being NGC~6034 and UGC~10143A (a.\ k.\ a.\ 16000+1606), 
respectively. (Note: their coordinates for NGC~6034 are incorrect; the correct 
coordinates are 16$^{\rm h}$01$^{\rm m}$16.4$^{\rm s}$,
17\arcdeg 20\arcmin 07\arcsec (B1950).) These claims are apparently 
based on the surface photometry of Oemler (1976) and  Schombert (1986), 
who noted that the surface brightness profiles of these galaxies show  
the extended envelope characteristic of cD galaxies. 
UGC~10143A is the brightest galaxy in A2147 and is located near the 
cluster center, presumably at the bottom of its potential well. 
NGC~6034 is not the brightest galaxy in A2151, and is located far from 
the cluster center. It is, however, located in the center of the 
SW group discussed in Section~\ref{vel-str}; this may be further evidence
that this group is a
dynamically distinct subclump of A2151. Even if NGC~6034 is
at the bottom of the group's potential, the `well' is not very deep.
NGC~6034 has unusual radio features: it dominates the radio continuum emission 
in this field and shows rare HI absorption (\cite{dic97}).
Huang \& Sarazin (1996) report that the brightest cluster galaxy in A2151, 
NGC~6041A, {\it is} located at the central X-ray brightness peak and the 
optical cluster center, but that ``it is certainly not a D or cD galaxy''. 
This apparent contradiction between the cluster types and the existence of
cD galaxies does not appear to have been mentioned before.

\section{Supercluster Dynamics}\label{sup-dyn}

To see if the mean velocity differences between the clusters corresponded to
physical separations, we attempted to determine the true line-of-sight 
position of the three clusters using several methods. We analyzed the 
distances given by Buta \& Corwin (1986) from use of the $B$ band 
Tully-Fisher relation. From the mean errors 
given for total magnitude and HI line width, we calculate the errors 
in their distances to be $\sim 10$\% per galaxy, which is rather small since
Pierce \& Tully (1992) estimate the scatter in the $B$ band Tully-Fisher relation
as $\sim 0.3$~magnitudes, or $\sim 17$\% in distance. Since the errors in
individual distances are likely to be large, we calculated distances to the 
clusters by averaging the distances of all galaxies 
in each cluster. (We identified their galaxies with ours on the basis of 
position and velocity and used our cluster assignments.) 
Unfortunately, there were only a few galaxies with distance estimates
in each of A2147 and A2152.
We quote two sets of results: one based on the line width data from 
Giovanelli, Chincarini, \& Haynes (1981), marked `GCH', and one based on 
all other data, since Buta and Corwin remark that these 
two samples are different. Both show that the distance
of the clusters correlates with their redshift, although the absolute
distances are quite different. We found a similar ordering of distances from
calculating distances using the brightest cluster galaxy method and data of 
Lauer \& Postman (1994) and Postman \& Lauer
(1995). They estimate their typical distance accuracy as 17\% per BCG. 
A summary of all three sets of distances is in Table~\ref{tfdata}. Since 
the distance
errors may be large, we regard these distance results as mildly supportive of
our kinematical results but do not use them in further computations. 

One indicators of the dynamical state of the supercluster is the
crossing time compared to a Hubble time. We computed several crossing times
for the supercluster: the virial crossing time $\Delta t_v$, the moment of inertia 
crossing time $\Delta t_I$, and the linear moment crossing time $\Delta t_L$. 
We used the formulae of Gott \& Turner (1977): 
\begin{equation}
\Delta t_v=\frac {3 \pi}{10 \sqrt{5}}\frac{\bar{v}}{{\sigma}_v} \sin \phi, 
 \quad \phi=N\left({\sum}_{\rm pairs} 1/{\theta}_{ij}\right)^{-1}
\end{equation}
\begin{equation}
\Delta t_I=\frac{\bar{v}}{\sqrt{2}{\sigma}_v}\left(\frac{\sum_i {{\theta}_i}^2}
{N}\right)^{1/2}
\end{equation}
(${\theta}_i=$ angular distance from cluster center to supercluster
center of mass)
\begin{equation}
\Delta t_L=\frac{2}{\pi}\bar{v} \frac{\left<\sin {\theta}_{ij}\right>_{\rm pairs}}
{\left<v_i-v_j\right>_{\rm pairs}}
\end{equation}
where all sums are over the three clusters. All three crossing times 
are approximately 10\% of the Hubble time: $\Delta t_v H_0 =0.08$,  
$\Delta t_I H_0= 0.09$, and $\Delta t_L H_0 = 0.11$.   
This indicate that the supercluster is bound.

Another way to determine whether the supercluster is bound is to the 
Newtonian binding condition of Davis et al.\ (1995), derived from the
two-body models of Beers, Geller, \& Huchra (1982).
A pair of gravitating masses is bound if
\begin{equation}
{V_r}^2 R_p \leq 2GM {\sin}^2 \alpha \cos \alpha
\label{eq-bound}\end{equation}
where $\alpha$ is the angle between the plane of the sky and the 
true line separating the two clusters, $R_p$ is their projected separation, 
and $V_r$ is their relative line-of-sight velocity. The true physical and 
velocity separations are $V=V_r/\sin \alpha$ and $R=R_p/\cos \alpha$.
From the available information we can determine the range of possible $\alpha$
over which the three pairs of clusters could be bound.

Table~\ref{2body} shows the values of $V_r$ and $R_p$ for each of the three
pairs, and the ``binding ratio'' (the left-hand side of Eq.~\ref{eq-bound})
for all three. The results show that, for our calculated velocity differences, 
the A2147/A2151 system is bound for $13\arcdeg <\alpha<88\arcdeg$
A2152 would not be bound to A2151 or A2147 for any projection angle.
If we consider A2151 and A2147 to be a single system
located at the midpoint of their projected positions and radial velocities,
then A2152 would be bound to this system for $35\arcdeg <\alpha<73\arcdeg$.
Colless \& Dunn (1996) point out that the probability of observing a system
with projection angle ${\alpha}_1 < \alpha < {\alpha}_2 $ is proportional
to $\sin({\alpha}_2)-\sin({\alpha}_1)$; this gives probabilities of 
40\% that A2152 is bound to A2151+A2147, and 77\% that A2147 is bound to A2151.
From this we conclude that A2147 and A2151 are probably 
bound to each other, but that there is not good evidence for the supercluster
as a whole to be bound.

Tarenghi et al.\ (1980) made a limited attempt at studying the clusters as
part of a three-body system. They compared the sum of the virial 
masses of the three 
clusters to the virial mass of the entire supercluster complex (considering
all galaxies as individual members of the supercluster), and found the two
to be roughly comparable. They also computed the virial mass of the three
mass-point system comprised of the three clusters, but found a very low mass
($2\times10^{14}h^{-1}M_{\sun}$, less than their individual cluster masses.)
With an improved redshift sample, and a better 
separation of the galaxies into clusters, we can improve upon their work. 
The supercluster virial and projected masses, calculated assuming the 
three clusters as mass points located at their central positions
and radial velocities, and, of course, assuming the supercluster is relaxed,
are $(8.6 \pm 1.2) \times 10^{15} h^{-1} M_{\sun}$ and 
$(6.6\pm 1.0) \times 10^{15} h^{-1} M_{\sun}$.  
The sums of the cluster masses are $(2.7 \pm 0.3) \times 10^{15} h^{-1} M_{\sun}$ 
and $(3.8 \pm 0.4) \times 10^{15} h^{-1} M_{\sun}$. 
Since the two sums of cluster masses are comparable to the binding mass 
for the supercluster (one half of the virial mass), this can be
considered additional evidence
that the supercluster is bound. From all of the above results (crossing
times, binding ratios, and binding mass) we consider it reasonable to
assume the supercluster is marginally bound.

Assuming the supercluster to be bound means that it is reasonable to
calculate its mass-to-light ratio to derive a value for $\Omega$.
We computed the total luminosity of the supercluster using a similar method to 
that used in Section~\ref{sec-clpar}: computing the total luminosity from the
galaxy magnitudes and making a faint-end correction. The Schechter function 
used was one we fit to the data, with parameters ($M_*=-19.74$, $\alpha=-0.92$).
The total supercluster luminosity was $(1.4\pm0.2) \times 10^{13} h^{-2} L_{\sun}$. 
An upper limit for the volume of the supercluster (assuming 
it to cover 22.5 square degrees and $v=8500$~km~s$^{-1}$ to $v=14500$~km~s$^{-1}$),
is $5.6\times 10^3 h^{-3}$ Mpc$^3$, giving a lower limit to the luminosity density of 
$(2.4 \pm 0.3 ) \times 10^9 h$ Mpc$^{-3}$. This is an overdensity of a factor of 13 
compared to the LCRS field luminosity density. A plausible lower limit to the
supercluster volume is the volume enclosed in the virial radius $R_v=\frac{\pi}{2}R_h$
where $R_h$ is the harmonic radius (\cite{car96}).
The computed virial radius for the three clusters as mass points is 3.75 $h^{-1}$ Mpc, 
giving a volume of 222 $h^{-3}$Mpc$^3$. The corresponding luminosity 
density, $(3.3 \pm 0.6) \times 10^{10} h$ Mpc$^{-3}$, is an overdensity 
of a factor of 190 compared to the field.

The mass-to-light ratio of the supercluster can be computed
from the above dynamical parameters. The mass-to-light ratio of the supercluster,
using $M=(7.6 \pm 2.0) \times 10^{15} h^{-1} M_{\sun}$ (the average
of the virial and projected masses) is then $(530\pm 160) h (M/L)_{\sun}$;  
this ratio yields a value for $\Omega$ of $\sim0.34\pm 0.1$. This is very close
to the value $\Omega = 0.36$ derived by Small et al.\ (1998) for the Corona Borealis 
Supercluster, although they used a very different method (calculating the virial
mass considering all galaxies as individual members of the supercluster).
Postman, Geller, \& Huchra (1988) derived $\Omega=0.2\pm0.1$
for Cor Bor by assuming the supercluster had the same $M/L$ as the clusters; this
is also compatible with our result. 
 
\section{Conclusions} 
We have demonstrated that the assignment of galaxies to adjacent clusters
in a supercluster can have significant effect on the clusters' dynamical
parameters. We conclude that the KMM algorithm is a useful
tool for this cluster assignment procedure. We find that A2152 and A2147 were 
probably confused in previous studies, and that the velocity dispersions of both are 
lower than those in previous work (715 and 821 km~s$^{-1}$, respectively); 
further, A2152 has a slightly larger mean velocity.
Distance measurements of the clusters support this assessment. 
Our dynamical measurements of the supercluster support the conclusion
that it is bound; its mass-to-light ratio yields a value for $\Omega$
of $0.34\pm0.1$, compatible with other measurements from superclusters.

\acknowledgements

We thank L. Macri and W. Brown for obtaining CCD images of the supercluster,
P. Berlind, J. Peters, and P. Challis for assistance with obtaining redshifts,
and D. Fabricant for building the FAST spectrograph.
We thank the referee for helpful suggestions, J.M. Dickey for 
communicating new redshifts in advance of publication, 
and C. Bird for providing us with the ROSTAT and KMM software.
This research has made use of the APS Catalog of the POSS I
which is supported by the National Science Foundation, the National Aeronautics
and Space Administration, and the University of Minnesota.  The APS
databases can be accessed at {\it http://isis.spa.umn.edu/}.
The Digitized Sky Survey was produced at the Space Telescope Science Institute
under US Government grant NAG W-2166. The images of these surveys are based on 
photographic data obtained using the Oschin Schmidt telescope on Palomar Mountain
and the UK Schmidt telescope.  The plates were processed into the present 
compressed digital form with the permission of these institutions. 
This research has made use of the NASA Astrophysics Data System Catalog Service.

\appendix

\section{Catalog construction}
The ideal catalog for a dynamical study of a supercluster is one that is 
magnitude-limited, so that a spatially uniform sample of galaxies can be made.
While several catalogs of galaxies in Hercules or A2151 have been made
(\cite{dic87}, \cite{dre88}), none covered all three clusters and hence are
not ideal for our study. The original Palomar Sky Survey can be used in two ways
to generate catalogs of galaxies in nearby clusters: using automatic detection
techniques such as FOCAS on images from the Digitized Sky Survey (\cite{las91}), 
and extracting similar information from a publicly available catalog of the 
Minnesota Automated Plate Survey. 
We used both techniques to construct a catalog of galaxies in Hercules,
and our intentions were both to test them against each other and to use them
to complement each other. 
Our goal was to create as complete a galaxy catalog as possible,
with a magnitude limit such that it would include at least 200 galaxies in Hercules
without measured redshifts.

\subsection{APS}
The Minnesota Automated Plate Survey (\cite{pen93}, hereafter referred to 
as APS) is a catalog of all the objects detected in digitizing scans of 
the O (blue) and E (red) plates of the original Palomar Sky Survey. 
The objects' coordinates, 
magnitudes, sizes, shapes, and classifications are available. The 
classification, done by a neural network algorithm, is given in the form 
of the probability that the object is a galaxy (the parameter {\it node\_gal});
an object is assumed to be a star if it is not a galaxy.  
The catalog is accessible over the World Wide Web through the NASA 
Astrophysics Data System.

To make a list of objects in the supercluster, we made a query for all objects
in our region with {\it node\_gal(O)} $>0.5$ and $m_{\rm APS}(E) <16.5$. 
We used the O plate classifications because, according to the APS catalog 
documentation, 
they are more accurate. We used a magnitude limit from the E plates to more closely 
match
the results of the FOCAS catalog, which is also based on the E plates.
The low limit of {\it node\_gal} was chosen so that as few galaxies as possible 
would be missed.
Unfortunately, as seen below, this resulted in a large amount of contamination
of the galaxy list by stars. Since we did not know the
transformation of APS magnitudes to a standard system, we chose 
the limit $m_{\rm APS} (E) <16.5$ so that the resulting catalog would have a
reasonable number of objects; our APS catalog contained 1142 galaxy candidates. 

\subsection{FOCAS}
We also used the `faint object classification and analysis system' 
in IRAF (\cite{val82}) to extract a list of galaxies in the region. 
FOCAS detects, measures (areas, moments, and magnitudes), and classifies
all of the objects in an image. The classification algorithm 
fits templates based on the PSF to the objects; objects are classified
based on the parameters of the best-fitting template. 
The result of running FOCAS is a catalog with entries similar to 
those of the APS\@. FOCAS attempts to split multiple or overlapping objects 
into components; this task is not always successful, since it tends to split 
bright galaxies into multiple parts unnecessarily. This was not a problem 
for the fainter galaxies but introduces an additional source of error into the 
magnitudes of bright objects. We ran FOCAS using the standard script `autofocas' 
described in the 
documentation (\cite{val82}) and the suggested input parameters: ${\rm Nmin}=20$,
${\rm sigma}=0.1$, ${\rm size}=10$. The saturation level was set at 15100 counts
to provide an adequate discrimination between stars and galaxies. The catalog
magnitude limit was set at 21.5; all other parameters were left at the 
FOCAS defaults, including the classification rules.

We constructed a FOCAS catalog from the Digitized Sky Survey (\cite {las91})
image of the region, made from the Palomar Sky Survey E plates. 
Using the FOCAS-provided pixel centers and the transformation routines 
provided in the DSS documentation, we computed coordinates for each object.
No photometric information was available 
for the Digitized Sky Survey images, so the FOCAS magnitudes were determined
assuming a linear pixel-values-to-intensity relation and the default
zero point: $m_{\rm FOCAS} = 30.0-2.5\log({\rm intensity})$. We used the
FOCAS total magnitudes, which sums the intensity inside the FOCAS `total area', 
as these have been shown to be less biased than isophotal magnitudes (\cite{wei95}).
(We checked the linearity assumption by comparing star profiles made from the DSS 
and from a CCD image; the DSS pixel values (photographic density)
were linear with intensity, with a correlation coefficient $r^2=0.98$.)

We extracted a list of FOCAS-classified `galaxies' from the catalog, 
with (again, an arbitrary) magnitude limit of $m_{\rm FOCAS}=16.9$.  
To the FOCAS `galaxy' list we added a list of `potential
galaxies' which FOCAS had classified as stars. We suspected that these objects
might be galaxies because of large ellipticity or area, but they had been automatically 
classified as stars because they were saturated. (Changing the FOCAS saturation level
parameter so that these objects were classified as galaxies was found to result
in a very large contamination of the `galaxy' list by stars.) The resulting 
FOCAS list of `galaxies' and `potential galaxies' had 591 entries. We also 
constructed another `larger' FOCAS catalog, containing all objects to the same magnitude
limit regardless of classification; there were 10043 objects in this list.  

We constructed a catalog of galaxies with measured velocities using NED\footnote{The 
NASA/IPAC Extragalactic Database is operated by the Jet Propulsion
Laboratory, California Institute of Technology, under contract from the National 
Aeronautics and Space Administration.} and ZCAT (\cite{huc96}).
NED was taken to be the auxiliary source, with all ZCAT objects
going into the reference catalog, and NED objects included only
if they were not already in ZCAT. Since the two catalogs use different
naming conventions, we merged them using
a matching procedure: two galaxies were considered to be the same object if
they were within a specified distance both in position and in velocity.
In practice, the velocity information was more useful. Velocities 
for the `same' object (objects were assumed to be the same if 5\arcmin \ diameter
fields centered on each of their coordinates contained the same brightest galaxy)
were generally within 100~km~s$^{-1}$, even if the catalog coordinates were 
disparate by several arcminutes. To this reference catalog list we added a 
list of galaxies whose velocities were made available as part of a separate 
infrared-selected survey in the Hercules region (\cite{hu97b}). 

The catalog of galaxies whose redshifts would be measured was constructed by 
first merging the APS and FOCAS lists. This procedure showed that the astrometric
calibrations of both the APS and DSS were excellent: `matched' objects typically
differed in position by only a few arcseconds. Some objects in both the APS and FOCAS
lists were not matched with objects in the opposite list. By searching the full APS
catalog and the FOCAS `larger' catalog, we found that all of these `unmatched'
objects were included in the catalog they were `missing' from, but were
classified as something other than a galaxy. In particular, FOCAS often classified
galaxies as type `d' (for `diffuse'). We found the FOCAS equivalent
object, and its magnitude, for each unmatched APS object, so that the merged
list would have a uniform source of magnitudes. We then removed galaxies in the 
reference catalog (which already had measured velocities) from the merged list. 
All of the reference catalog galaxies had APS and FOCAS counterparts, once
position errors, misclassification, and the magnitude limit had been accounted for. 
We estimated the completeness of our merged list by comparing it with the complete
list of Zwicky galaxies in the region, to a B magnitude of 15.5. 122 of these 132 
galaxies were in the merged list and classified as galaxies; all of the remaining 10
were classified as stars in the full APS and `larger' FOCAS catalogs. 4 of these
10 galaxies were bright NGC/IC objects (which would not likely have required 
velocity measurements), 
so our completeness is $\sim 126/132=95\%$. 

The 958 remaining objects in the merged list were examined visually on the DSS
images to determine if they were in fact galaxies. To separate galaxies from
merged stars we looked for the existence of a bulge or disk in a object, and compared
the elongation and size of the objects to the PSF of nearby stars. This last
is important because aberrations mean that the PSF is not round near the
edge of a plate.  The overall `galaxy yield' for the merged list was 31\%;  
the yield was higher for objects classified as galaxies by both FOCAS and APS  
(55\%), and lower for objects classified as galaxies by FOCAS alone (13\%), 
or APS alone (25\%). Objects misclassified as galaxies were most
often two nearby or overlapping stars (45\%), single stars with diffraction 
spikes (18\%), or overlapping faint galaxies (4\%). This result is not unexpected 
since the 
surface density of stars is about fifty times that of galaxies at this magnitude
but implies that stars misclassified as galaxies can severely contaminate a 
`galaxy' list
which is not carefully examined. The FOCAS `double stars' were usually
closer than the APS `double stars' (the images touched or were less than a PSF
diameter apart), which implies that FOCAS is better at separating nearby objects
than the APS algorithm. 

\begin{figure}
\plotone{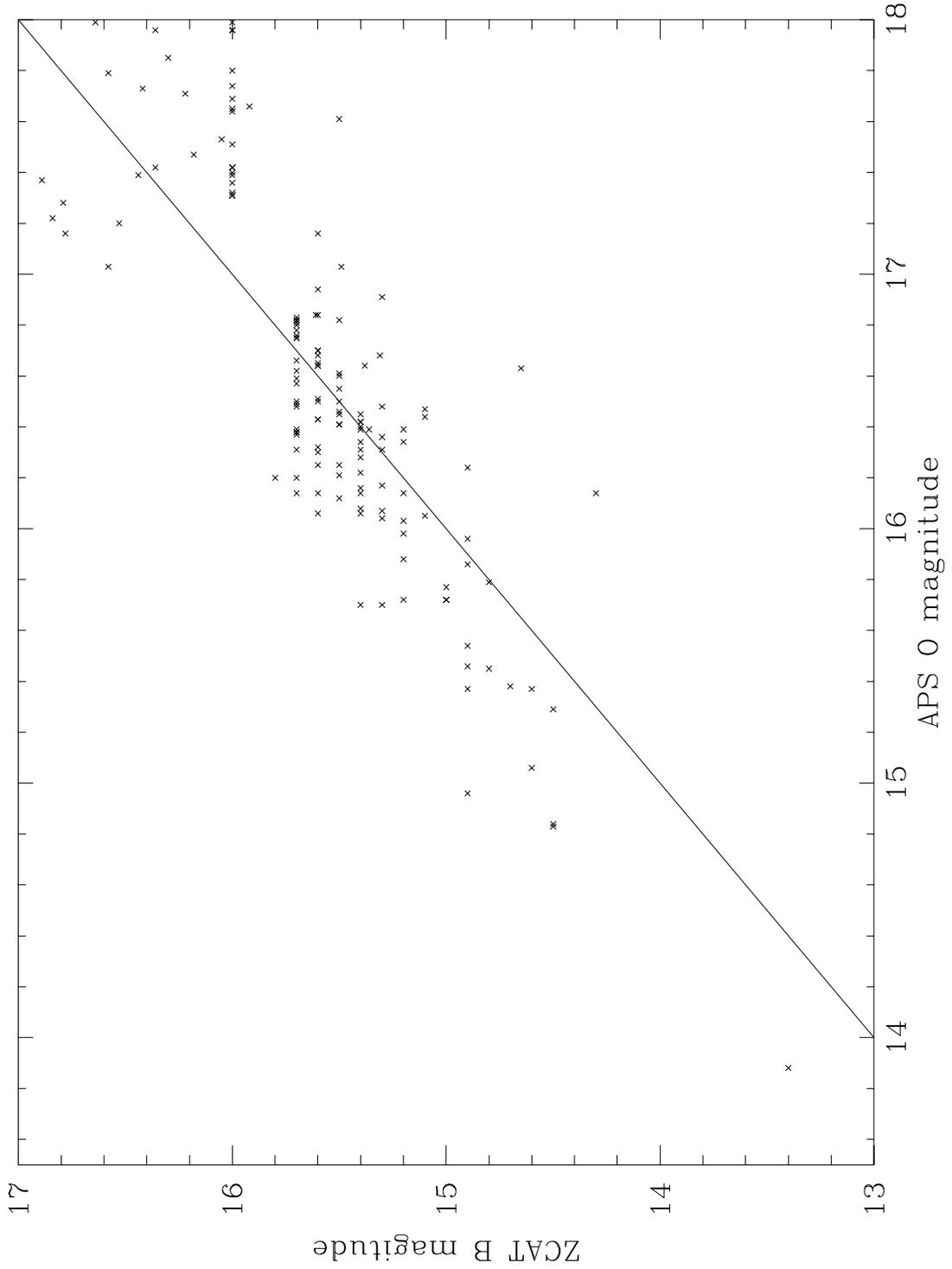}
\caption{Rough calibration for APS and ZCAT magnitudes: line is
  m$_{Zw}$=m$_{\rm APS}$-0.93\label{aps-z-cal}}
\end{figure}

The resulting list of galaxies (293 in total) were magnitude-ordered using the 
FOCAS total magnitudes; redshifts were measured for the brightest of these galaxies.
For the sake of interest, we computed the transformation between APS O
magnitudes and the reference catalog magnitudes
(which were on the Zwicky $B(0)$ system (\cite{huc76}))
during the removal of reference catalog galaxies from the merged list. 
We found that the APS O magnitudes were fainter than the 
catalog magnitudes ($m_{\rm APS}(O)-m_{\rm Zw}=+0.93\pm 0.05$, 
see Figure~\ref{aps-z-cal}). 
Some of the features of Figure~\ref{aps-z-cal} are due to artifacts of the reference
catalog compilation process; specifically, $B=15.7$ is the Zwicky catalog limit,
and the large number of galaxies at $B=16$ is due to imprecise magnitude estimates in 
the Uppsala General Catalog. Given the APS catalog construction procedures, this large
magnitude offset is not unexpected (\cite{cab96}). 
The scatter of about 0.3 magnitudes in the offset is also not unexpected
given the 0.3 magnitude scatter in the Zwicky magnitudes (\cite{huc76}; 
\cite{bot90}).

\begin{figure}
\plotone{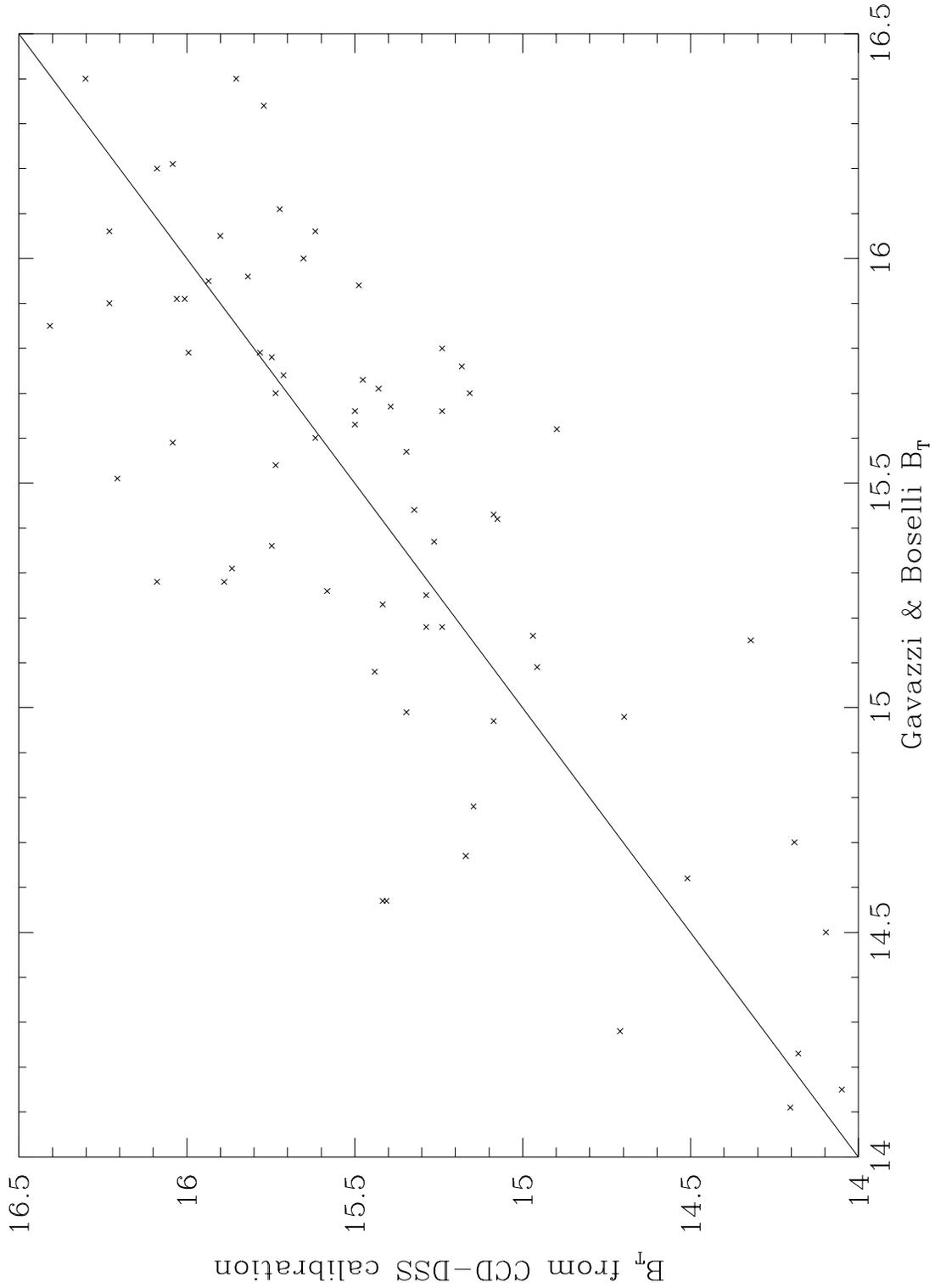}
\caption{Comparison of $B$ magnitudes from CCD-DSS calibration to
Gavazzi \& Boselli $B_T$ magnitudes. Solid line has slope 1, intercept 0.
\label{gav-comp}}
\end{figure}

\begin{figure}
\plotone{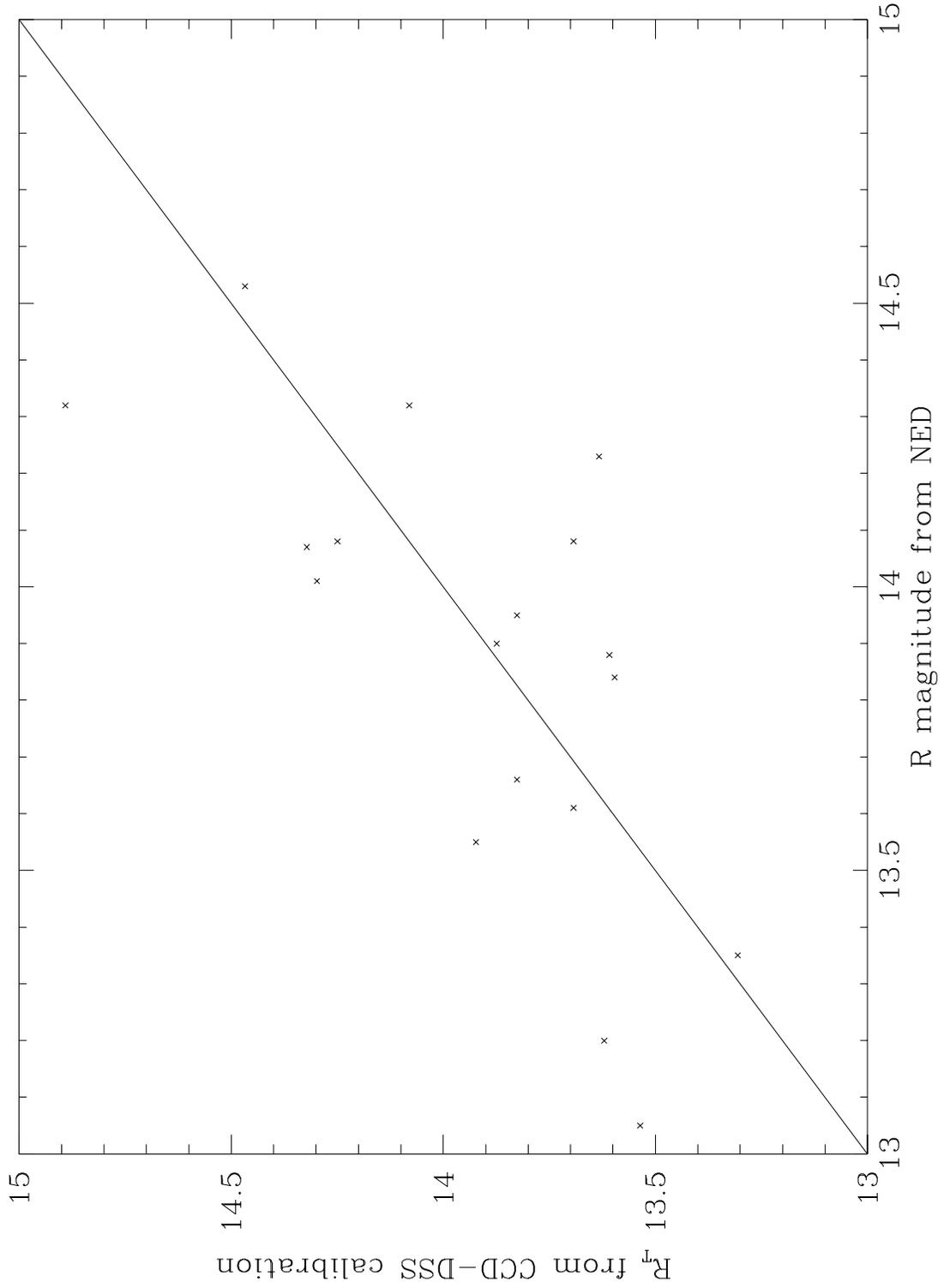}
\caption{Comparison of $R$ magnitudes from CCD-DSS calibration to
NED $R$ magnitudes. Solid line has slope 1, intercept 0.
\label{ned-comp}}
\end{figure}

\begin{figure}
\plotone{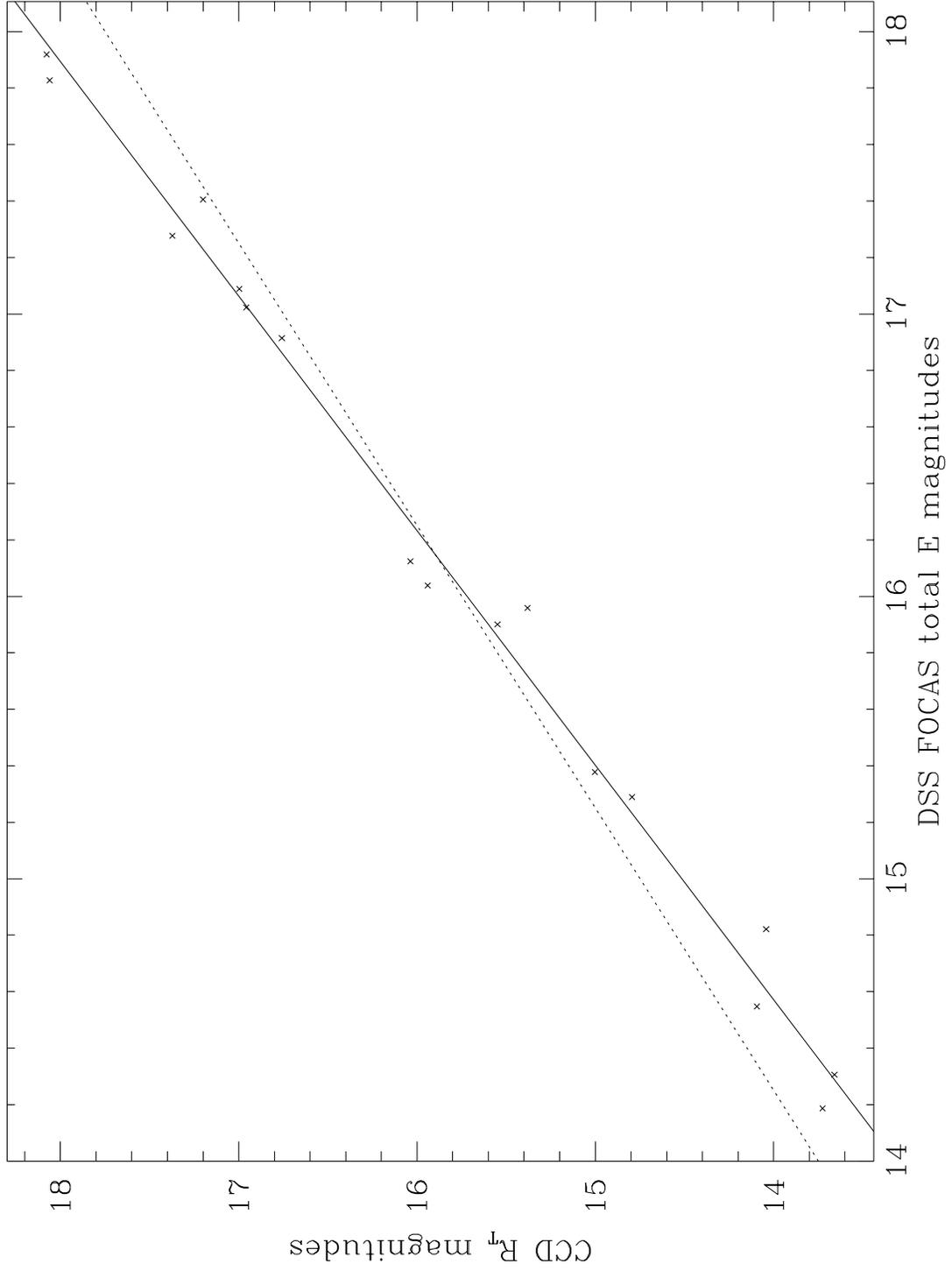}
\caption{Calibration for CCD $R_T$ and DSS $E$ magnitudes. Solid line: least squares fit.
Dashed line: least squares fit with slope forced to 1.
\label{focas-cal}}
\end{figure}

In order to have a uniform set of magnitudes for all galaxies in the supercluster, 
we determined the $R$ magnitudes from the FOCAS total magnitudes measured on the DSS\@. 
To calibrate the FOCAS magnitudes, we obtained CCD images (on
photometric nights) in $B$ and $R$ of several 
fields in the region using the 1.2m telescope at FLWO\@.
We reduced the CCD images in the standard manner, and
measured asymptotic total magnitudes of galaxies using a series of apertures
(see Table~\ref{phot-2151}).
We did a least squares fit of both $B$ and $R$ magnitudes against the 
DSS magnitudes in order to compare our photometry to
the $B_T$ magnitudes published by Gavazzi \& Boselli (1996)
and available $R$ magnitudes from NED. The results are in Figures~\ref{gav-comp}
and \ref{ned-comp}; there is no evidence of a significant scale error
or zero-point shift in either color.
(The large scatter in the $B$ magnitude plot is to be expected since
we ignored any color term present in the transformation from the red DSS 
magnitudes to $B$ magnitudes.)
The least-squares fit for the calibration relation was $R=1.22 m_{\rm FOCAS} - 3.55$, 
with a scatter of $\sim$ 0.14 magnitudes and scale error of 0.18 magnitudes/magnitude. 
This is a fairly large scale error; however, the relation given clearly fits the 
data better than a least-squares fit with the slope forced to 1 
(see Figure~\ref{focas-cal}). 
This relation and the FOCAS total magnitudes were used to 
derive $R$ magnitudes for all of the galaxies in our list; we report these in
Table~\ref{all-dat} only to the nearest 0.1mag due to the large scatter of the fit.

\clearpage
\begin{deluxetable}{lllrrrlc}
\tablecaption{Galaxies in the Hercules supercluster\label{all-dat}} 
\tablehead{
\colhead{Name} &  \colhead{$\alpha $ (J2000)} & \colhead{$\delta $ (J2000)} & \colhead{$R$}
& \colhead{$v$}  & \colhead{$\Delta v$} & \colhead{type} & \colhead{velocity source}}
\startdata
           15535+1826 &  15 55 43.07 & 	18 16 57.29 & 14.8 &  5327 &  22 & S &* \nl
           15541+1640 &  15 56 23.71 & 	16 31 19.99 & 13.1 &  4630 &  71 & E &(21)  \nl 
           1554+1847B &  15 56 56.00 & 	18 38 20.00 & 15.1 & 18138 & 100 & E &(13)  \nl 
           1554+1847A &  15 56 58.99 & 	18 38 34.98 & 14.1 & 18138 & 100 & S &(13)  \nl 
           15548+1746 &  15 57 04.25 & 	17 37 31.80 & 14.7 & 11086 &  40 & S &*     	\nl
           15548+1819 &  15 57 04.39 & 	18 11 12.01 & 14.3 &  9545 & 100 & E &(24)  \nl 
                N6018 &  15 57 29.84 & 	15 52 22.12 & 12.8 &  5218 &  25 & S &(17)  \nl 
                N6021 &  15 57 30.74 & 	15 57 21.17 & 12.7 &  4738 &  25 & E &(17)  \nl 
           15553+1810 &  15 57 35.39 & 	18 01 32.55 & 14.3 &  9429 &  20 & S &(6)  \nl 
           15553+1617 &  15 57 36.76 & 	16 08 02.22 & 14.6 & 10667 &  26 & S &*     	\nl
           15554+1820 &  15 57 40.39 & 	18 11 17.99 & 15.1 & 14433 &  39 & S &*     	\nl
           15554+1616 &  15 57 40.50 & 	16 07 30.14 & 15.6 & 11173 &  21 & S &*     	\nl
           15554+1621 &  15 57 42.91 & 	16 13 03.00 & 14.9 & 10120 & 100 & S &(13)  \nl 
           15555+1631 &  15 57 46.47 &  16 22 24.24 & 15.1 & 10850 &  24 & S &*  \nl
                N6022 &  15 57 47.70 & 	16 16 56.24 & 14.8 & 11225 & 100 & S &(24)  \nl 
                N6023 &  15 57 49.64 & 	16 18 35.39 & 12.7 & 11140 & 150 & E &(1)  \nl 
          CGCG108-023 &  15 57 51.99 & 	16 21 40.00 & 15.4 & 13500 &  51 & S &(2)  \nl 
           15557+1505 &  15 58 01.99 & 	14 57 40.00 & 14.7 & 11275 & 100 & S &(25)  \nl 
           15557+1629 &  15 58 03.00 & 	16 20 44.99 & 13.8 & 10853 &  20 & S &(7)  \nl 
           15558+1528 &  15 58 04.48 & 	15 18 59.47 & 14.5 & 11247 &  44 & E &*     	\nl
           15559+1815 &  15 58 06.78 & 	18 06 51.23 & 14.7 & 13668 &  46 & E &*     	\nl
           15559+1745 &  15 58 14.45 &  17 36 10.08 & 14.9 & 11135 &  24 & S &*\nl
           15560+1629 &  15 58 18.26 & 	16 20 18.10 & 15.4 & 11485 &  42 & S &*     	\nl
           15561+1813 &  15 58 20.49 & 	18 04 50.88 & 14.2 & 13668 &  48 & S &*     	\nl
          CGCG108-027 &  15 58 26.22 & 	18 02 21.73 & 14.3 & 12642 &  39 & S &(28)\nl 
                I1151 &  15 58 31.98 & 	17 26 35.05 & 12.8 &  2169 &	5 & S &(9)  \nl 
           15563+1801 &  15 58 32.41 & 	17 52 17.04 & 14.4 & 14345 &  41 & S &*     	\nl
           A1556+1712 &  15 58 35.98 & 	17 04 17.29 & 15.1 & 13115 &  32 & E &(22)  \nl 
           15566+1920 &  15 58 47.25 &  19 11 41.57 & 14.8 &  8717 &  31 & S &*\nl   
           A1556+1808 &  15 58 47.71 & 	17 59 15.00 & 15.5 & 13445 & 410 & E &(22)  \nl 
           15565+1505 &  15 58 49.58 & 	14 58 05.09 & 14.3 & 10690 &  71 & E &(21)  \nl 
           A1556+1712 &  15 58 52.50 & 	17 03 50.01 & 15.3 & 17681 & 113 & E &(22)  \nl 
           15566+1506 &  15 58 54.05 & 	14 58 53.36 & 14.3 & 10527 & 100 & S &(13)  \nl 
           15568+1503 &  15 59 04.99 & 	14 55 35.00 & 14.1 & 12711 &  20 & S &(7)  \nl 
           15570+1518 &  15 59 15.72 & 	15 10 35.72 & 14.2 & 12774 &  71 & S &(21)  \nl 
           15571+1851 &  15 59 19.43 & 	18 42 03.60 & 14.9 & 13927 &  40 & S &*     	\nl
           A1557+1517 &  15 59 21.08 & 	15 08 20.00 & 15.7 & 12840 & 100 & S &(22)  \nl 
           15573+1807 &  15 59 34.19 & 	17 58 11.75 & 15.0 & 11028 &  25 & E &*     	\nl
           15574+1921 &  15 59 36.24 & 	19 12 52.02 & 15.1 & 10578 &  35 & E &*     	\nl
           A1557+1852 &  15 59 44.20 & 	18 43 59.99 & 16.7 &  8972 & 100 & E &(14)  \nl 
           15575+1856 &  15 59 45.71 & 	18 48 02.02 & 13.1 &  8961 & 100 & S &(24)  \nl 
           15576+1544 &  15 59 58.67 & 	15 35 28.46 & 14.0 & 10099 &  10 & S &(8)  \nl 
           15579+1827 &  16 00 07.87 &  18 19 00.81 & 14.6 & 10755 &  31 & S &*  \nl
           15579+1831 &  16 00 09.58 &  18 23 03.12 & 14.8 & 18428 &  33 & E &*\nl
           15580+1831 &  16 00 14.76 & 	18 22 33.85 & 14.8 & 18094 &  40 & S &*     	\nl
          15580+1617S &  16 00 15.99 & 	16 08 34.98 & 17.1 &  9990 & 100 & S &(13)  \nl 
          15580+1617N &  16 00 16.71 & 	16 08 39.98 & 15.5 &  9914 & 100 & S &(13)  \nl 
           15580+1554 &  16 00 17.75 & 	15 45 23.65 & 14.4 &  4782 & 100 & E &(13)  \nl 
           15581+1935 &  16 00 20.95 & 	19 26 25.58 & 14.6 & 13827 &  28 & E &*     	\nl
           15582+1646 &  16 00 26.60 & 	16 37 36.23 & 15.3 & 10300 &  20 & E &(6)  \nl 
           A1558+1810 &  16 00 26.79 & 	18 02 07.01 & 14.9 & 13913 & 233 & S &(22)  \nl 
           15582+1824 &  16 00 27.80 &  18 15 49.29 & 15.1 & 10683 &  27 & S &*  \nl
           15583+1900 &  16 00 29.06 &  18 51 07.38 & 14.9 & 10671 &  26 & E &* \nl
                I1155 &  16 00 35.71 & 	15 41 07.80 & 13.7 & 10629 &  15 & S &(20) \nl 
           15584+1649 &  16 00 37.73 & 	16 40 12.93 & 14.8 & 10586 &  38 & S &*     	\nl
           15584+1651 &  16 00 43.20 & 	16 42 55.01 & 14.6 & 10653 &  20 & E &(6)  \nl 
           15585+1626 &  16 00 44.13 & 	16 17 08.12 & 14.7 & 10216 &  28 & S &*     	\nl
           15581+1813 &  16 00 47.30 & 	18 04 40.01 & 14.4 & 13155 & 100 & S &(25)  \nl 
           15586+1904 &  16 00 49.21 & 	18 55 43.00 & 16.7 &  9437 & 100 & E &(13)  \nl 
           15585+1517 &  16 00 51.48 & 	15 09 04.75 & 14.2 & 10156 &  20 & S &(6)  \nl 
           15586+1741 &  16 00 52.27 & 	17 32 43.87 & 13.9 & 13349 & 100 & S &(13)  \nl 
          15586+1628S &  16 00 53.71 & 	16 20 08.02 & 13.9 & 12382 &  71 & E &(21)  \nl 
           A1558+1631 &  16 00 54.40 & 	16 20 44.02 & 16.3 & 43000 & 100 & S &(13)  \nl 
          15586+1628N &  16 00 54.79 & 	16 20 42.00 & 14.7 & 13109 &  52 & S &(21)  \nl 
           15586+1540 &  16 00 56.40 & 	15 31 33.60 & 14.1 & 11567 & 104 & S &*     	\nl
           15587+1845 &  16 00 56.98 & 	18 36 52.09 & 14.3 & 10477 &  38 & E &*     	\nl
           15587+1534 &  16 01 01.52 & 	15 25 11.78 & 14.2 & 10323 &  75 & S &*     	\nl
                I1160 &  16 01 02.50 & 	15 29 40.45 & 14.9 & 10970 & 100 & S &(13)  \nl 
           15589+1655 &  16 01 10.32 &  16 47 08.20 & 14.9 & 10638 &  43 & E &* \nl
           15589+1554 &  16 01 12.36 &  15 45 21.78 & 15.0 & 11369 &  52 & S &*  \nl
           15590+1550 &  16 01 14.45 & 	15 41 19.21 & 14.8 & 10750 &  28 & S &*     	\nl
                I1162 &  16 01 16.29 & 	17 40 40.41 & 14.5 & 13273 & 100 & S &(13)  \nl 
           15590+1754 &  16 01 16.50 & 	17 46 00.01 & 14.5 & 10378 & 100 & S &(13)  \nl 
                I1161 &  16 01 16.90 & 	15 38 39.98 & 14.1 & 10852 & 100 & E &(13)  \nl 
           15591+1704 &  16 01 19.34 & 	16 55 49.04 & 14.9 & 10801 &  49 & S &*     	\nl
           15590+1627 &  16 01 21.29 & 	16 18 20.01 & 13.5 & 11282 & 100 & S &(13)  \nl 
           15591+1649 &  16 01 21.68 & 	16 40 30.00 & 13.8 &  9473 &  71 & S &(21)  \nl 
           15591+1621 &  16 01 23.13 & 	16 13 04.80 & 14.7 &  8566 & 100 & S &(21)  \nl 
           15592+1914 &  16 01 25.71 & 	19 06 05.80 & 15.1 & 12431 &  44 & E &*     	\nl
           15592+1704 &  16 01 27.30 & 	16 55 10.96 & 14.7 & 10464 &  33 & S &*     	\nl
           15592+1723 &  16 01 27.99 & 	17 14 21.99 & 14.2 & 10765 & 100 & E &(13)  \nl 
           A1559+1809 &  16 01 28.81 & 	18 01 04.22 & 14.5 & 11053 &  89 & S &(22)  \nl 
                N6028 &  16 01 28.89 & 	19 20 30.01 & 13.0 &  4475 &  20 & S &(13)  \nl 
           15592+1653 &  16 01 30.00 & 	16 45 20.23 & 14.3 &  9598 & 100 & E &(13)  \nl 
                I1163 &  16 01 30.47 & 	15 30 14.25 & 13.7 & 10503 &  71 & E &(21)  \nl 
           15592+1558 &  16 01 32.02 & 	15 49 49.94 & 15.0 & 11474 &  23 & E &*     	\nl
           15593+1632 &  16 01 32.70 & 	16 23 33.93 & 15.3 &  8753 &  32 & E &*     	\nl
          A1559+1645A &  16 01 36.37 & 	16 36 40.64 & 18.1 &  4904 & 233 & S &(22)  \nl 
          A1559+1645B &  16 01 36.37 & 	16 36 40.64 & 17.4 & 12669 & 194 & E &(22)  \nl 
           15593+1634 &  16 01 36.77 & 	16 25 52.68 & 14.3 & 12727 &  20 & S &(6)  \nl 
           15594+1627 &  16 01 39.43 & 	16 18 36.97 & 15.0 &  8704 &  23 & S &*     	\nl
           15594+1805 &  16 01 41.36 &  17 57 03.31 & 15.1 &  4242 &  56 & S &*  \nl
           15595+1558 &  16 01 46.88 & 	15 49 19.52 & 15.1 & 11551 &  27 & S &*     	\nl
           15596+1853 &  16 01 49.29 & 	18 43 14.56 & 16.9 &  2627 &  40 & S &(10) \nl 
           A1559+1746 &  16 01 49.62 & 	17 38 26.52 & 15.1 & 10736 &  20 & S &(12) \nl 
           15596+1808 &  16 01 50.91 & 	17 59 43.08 & 14.9 & 14198 &  64 & S &*     	\nl
                N6030 &  16 01 51.46 & 	17 57 25.64 & 12.6 &  4491 &  25 & E &(16) \nl 
           15596+1556 &  16 01 51.96 & 	15 47 32.68 & 14.5 & 12406 &  46 & S &*     	\nl
          15597+1635N &  16 01 54.91 & 	16 27 15.01 & 14.5 & 10589 & 100 & E &(13)  \nl 
          KUG1559+158 &  16 01 55.83 & 	15 42 28.83 & 15.2 & 10423 & 100 & S &(31)  \nl 
            1559+1634 &  16 02 01.75 & 	16 26 07.26 & 14.9 &  9120 &  70 & E &(4)  \nl 
          15597+1635S &  16 02 01.89 & 	16 27 05.00 & 14.9 &  9112 & 100 & E &(13)  \nl 
           A1559+1653 &  16 02 01.89 & 	16 45 24.99 & 16.0 &  9662 & 122 & E &(22)  \nl 
           15598+1857 &  16 02 02.08 & 	18 49 00.33 & 15.5 &  2520 &  45 & E &(3)  \nl 
           15598+1713 &  16 02 04.23 & 	17 04 33.42 & 13.6 & 11036 &  71 & E &(21)  \nl 
               I1165N &  16 02 07.98 & 	15 41 44.99 & 15.6 & 14898 &  96 & S &(2)  \nl 
               I1165S &  16 02 08.49 & 	15 41 31.99 & 16.3 & 10122 &  59 & S &(2)  \nl 
           15599+1602 &  16 02 09.13 & 	15 53 16.84 & 14.8 &  9027 &  54 & S &*     	\nl
           15599+1657 &  16 02 12.67 & 	16 48 27.36 & 15.9 & 10044 &  55 & S &*     	\nl
          16000+1606S &  16 02 12.69 & 	15 54 26.93 & 14.3 & 10489 &  36 & E &(1)  \nl 
           15599+1634 &  16 02 12.95 & 	16 25 33.96 & 14.4 &  9276 & 100 & S &(13)  \nl 
          16000+1606N &  16 02 13.16 & 	15 56 14.97 & 14.3 & 13178 &  39 & S &(1)  \nl 
           1600+1609B &  16 02 14.00 & 	16 01 11.03 & 15.7 & 10256 & 100 & E &(13)  \nl 
               47-166 &  16 02 16.19 & 	16 04 40.84 & 20.8 &  9849 & 100 & S &(31)  \nl 
           16000+1606 &  16 02 16.94 & 	15 58 29.21 & 12.8 & 10384 &  36 & E &(1)  \nl 
           1600+1609A &  16 02 17.56 & 	16 00 10.22 & 15.7 & 10121 & 100 & E &(13)  \nl 
           16000+1630 &  16 02 18.02 & 	16 21 58.28 & 14.8 & 12054 & 232 & S &(21)  \nl 
           16000+1629 &  16 02 19.82 & 	16 20 44.37 & 12.6 & 11449 & 232 & E &(21)  \nl 
           16001+1536 &  16 02 21.06 & 	15 27 52.52 & 14.9 & 10424 &  32 & S &*     	\nl
           16001+1601 &  16 02 21.10 & 	15 52 41.05 & 15.6 & 10483 &  28 & S &*     	\nl
            1600+1451 &  16 02 21.30 & 	14 42 40.00 & 15.0 & 10812 & 100 & S &(27) \nl 
           16000+1617 &  16 02 21.52 & 	16 09 32.51 & 14.7 & 12894 & 232 & S &(21)  \nl 
           16001+1816 &  16 02 22.63 & 	18 07 44.22 & 15.1 & 18572 &  43 & S &*     	\nl
            1600+1824 &  16 02 24.47 & 	18 15 58.75 & 14.3 & 13682 &  50 & E &(23) \nl 
           16002+1632 &  16 02 29.67 & 	16 24 10.40 & 14.5 & 11507 &  27 & S &*     	\nl
           16003+2019 &  16 02 30.40 & 	20 10 34.71 & 14.8 & 25575 &  44 & S &*     	\nl
           16002+1442 &  16 02 32.82 & 	14 33 17.93 & 14.6 & 10241 &  28 & E &*     	\nl
               47-138 &  16 02 34.78 & 	15 44 59.82 & 17.4 & 11284 & 100 & S &(31)  \nl 
           16003+1637 &  16 02 36.17 & 	16 28 49.94 & 15.3 & 10543 &  47 & E &*     	\nl
           16005+1937 &  16 02 40.19 & 	19 29 12.48 & 15.6 & 12338 &  21 & E &*     	\nl
           16004+1809 &  16 02 40.22 & 	18 00 24.22 & 14.4 & 18498 &  33 & S &*     	\nl
           A1600+1553 &  16 02 40.49 & 	15 45 20.02 & 14.2 & 11020 & 251 & E &(22)  \nl 
           16004+1607 &  16 02 40.51 & 	15 58 59.81 & 15.8 & 10097 &  28 & E &*     	\nl
            1600+1649 &  16 02 43.30 & 	16 40 06.99 & 15.8 & 33338 & 100 & S &(13)  \nl 
           16005+1855 &  16 02 44.44 & 	18 46 54.55 & 15.4 &  9375 &  30 & S &*     	\nl
           16004+1508 &  16 02 44.55 & 	14 59 47.22 & 15.4 & 43036 &  68 & S &*     	\nl
           16005+1623 &  16 02 45.08 & 	16 14 53.59 & 14.8 & 10571 &  30 & E &*     	\nl
            1600+1525 &  16 02 45.67 & 	15 16 48.00 & 16.0 &  9252 & 100 & E &(14)  \nl 
           16005+1652 &  16 02 48.02 & 	16 43 23.20 & 15.1 & 10282 &  40 & E &*     	\nl
           16004+1615 &  16 02 48.55 & 	16 07 09.19 & 14.1 &  9933 & 100 & E &(13)  \nl 
            1600+1718 &  16 02 49.09 & 	17 10 05.27 & 14.5 & 10399 &  50 & S &(23) \nl 
            1600+1834 &  16 02 49.88 & 	18 26 40.38 & 15.6 & 12252 &  20 & S &(12) \nl 
           16006+1642 &  16 02 50.35 & 	16 34 11.32 & 13.7 & 10470 & 100 & E &(13)  \nl 
            1600+1605 &  16 02 50.93 & 	15 57 36.36 & 16.0 & 10465 &  70 & E &(4)  \nl 
           16006+1540 &  16 02 51.92 & 	15 31 48.79 & 15.0 & 17509 &  28 & E &*     	\nl
           16006+1528 &  16 02 54.66 & 	15 20 08.77 & 15.1 & 10095 &  21 & E &*     	\nl
           16007+1912 &  16 02 55.29 & 	19 03 48.92 & 16.5 & 13946 &  47 & S &*     	\nl
           16007+1905 &  16 02 57.40 & 	18 56 49.35 & 15.6 & 14013 &  28 & S &*     	\nl
           16007+1559 &  16 02 58.74 & 	15 50 33.83 & 13.8 &  9704 & 100 & S &(13)  \nl 
           16007+1615 &  16 03 00.56 & 	16 07 09.48 & 15.0 & 11974 &  32 & S &*     	\nl
           16007+1614 &  16 03 01.08 & 	16 05 45.67 & 14.9 & 11481 &  50 & S &*     	\nl
           16008+1536 &  16 03 03.67 & 	15 27 35.53 & 16.1 & 10428 &  40 & S &*     	\nl
           16008+1919 &  16 03 04.07 & 	19 10 55.99 & 15.1 & 14024 &  28 & S &*     	\nl
           16008+1624 &  16 03 05.28 & 	16 16 01.56 & 15.5 & 10963 &  29 & S &*     	\nl
       A2151:[D80]008 &  16 03 05.78 & 	17 10 14.44 & 16.4 & 10093 & 100 & S &(31)  \nl 
           16009+1617 &  16 03 09.82 & 	16 08 19.86 & 15.0 & 12460 &  31 & E &*     	\nl
           16009+1559 &  16 03 12.81 & 	15 50 54.53 & 14.9 & 11649 &  21 & E &*     	\nl
           16009+1502 &  16 03 13.48 & 	14 53 39.44 & 15.3 & 10907 &  31 & E &*     	\nl
            1600+1730 &  16 03 14.04 & 	17 22 02.86 & 14.8 & 10155 &  50 & E &(23) \nl 
           16010+1632 &  16 03 14.73 & 	16 24 09.87 & 13.8 & 10966 & 100 & S &(13)  \nl 
           16010+1913 &  16 03 15.87 & 	19 05 02.83 & 15.4 & 14111 &  46 & S &*     	\nl
            1601+1756 &  16 03 16.06 & 	17 47 47.01 & 14.8 & 11268 &  20 & E &(12) \nl 
           16011+1752 &  16 03 18.26 & 	17 43 57.14 & 15.5 & 10507 &  50 & E &*     	\nl
           A1601+1508 &  16 03 22.11 & 	15 00 24.98 & 16.6 & 16908 & 203 & S &(22)  \nl 
            1601+1711 &  16 03 23.01 & 	17 03 23.01 & 16.3 & 11666 &  50 & S &(23) \nl 
           16012+1918 &  16 03 26.71 & 	19 09 37.73 & 14.4 &  4684 &	8 & S &(11) \nl 
           A1601+1919 &  16 03 26.82 & 	19 10 47.75 & 15.8 & 11820 &  89 & E &* \nl 
            1601+1734 &  16 03 27.76 & 	17 25 58.73 & 16.5 & 11932 &  50 & S &(23) \nl 
                N6034 &  16 03 32.04 & 	17 11 55.00 & 13.5 & 10112 & 100 & E &(13)  \nl 
           16012+1542 &  16 03 32.26 & 	15 34 05.74 & 14.6 & 12399 &  44 & S &*     	\nl
            1601+1736 &  16 03 32.43 & 	17 28 47.03 & 14.8 & 13451 &  40 & S &(23) \nl 
           16012+1628 &  16 03 32.80 & 	16 19 23.01 & 13.3 & 11497 & 100 & E &(13)  \nl 
               47-030 &  16 03 35.14 & 	15 52 32.34 & 15.3 & 10147 & 100 & S &(31)  \nl 
           16013+1608 &  16 03 37.47 & 	15 59 23.10 & 15.5 & 11171 &  36 & E &*     	\nl
            1601+1602 &  16 03 38.02 & 	15 54 01.01 & 15.3 & 32821 & 201 & S &(13)  \nl 
           16014+1547 &  16 03 38.59 & 	15 38 43.83 & 15.0 & 11116 &  49 & S &*     	\nl
       A2151:[D80]006 &  16 03 38.95 & 	17 11 05.42 & 16.1 & 10217 &  45 & E &(23) \nl 
           16014+1459 &  16 03 39.71 &  14 50 58.78 & 14.5 & 11072 &  33 & S &* \nl
            1601+1728 &  16 03 40.46 & 	17 20 16.55 & 15.6 & 12873 &  50 & E &(23) \nl 
           16014+1637 &  16 03 41.61 & 	16 28 32.09 & 14.8 & 32239 &  29 & S &*     	\nl
           16014+1637 &  16 03 41.69 & 	16 28 30.07 & 14.8 & 11908 &  32 & S &*     	\nl
           16014+1605 &  16 03 42.78 & 	15 57 16.85 & 16.9 & 11935 &  25 & E &*     	\nl
          16014+1628W &  16 03 43.70 & 	16 19 35.00 & 14.3 & 10645 & 100 & S &(13)  \nl 
            1601+1555 &  16 03 43.81 & 	15 47 21.98 & 16.3 & 10185 & 200 & E &(15) \nl 
   DKP160130.51+18032 &  16 03 45.32 & 	17 55 11.86 & 15.9 & 33859 & 200 & S &(30)  \nl 
          16014+1628E &  16 03 45.79 & 	16 20 12.01 & 15.2 & 10645 & 100 & E &(13)  \nl 
           16015+1723 &  16 03 48.06 & 	17 14 26.02 & 13.8 & 10953 & 100 & S &(13)  \nl 
           16016+1605 &  16 03 50.15 & 	15 56 42.69 & 14.6 & 11059 &  47 & E &*     	\nl
           16016+1553 &  16 03 50.42 & 	15 44 52.83 & 15.0 &  9256 &  27 & S &*     	\nl
           16016+1505 &  16 03 52.85 & 	14 56 46.57 & 14.7 & 10261 &  34 & E &*     	\nl
           16016+1432 &  16 03 52.88 & 	14 23 49.35 & 15.0 & 10193 &  26 & S &*     	\nl
           16017+1913 &  16 03 53.74 & 	19 04 30.15 & 15.7 & 34032 &  37 & S &*     	\nl
           16016+1502 &  16 03 55.64 & 	14 54 09.03 & 14.4 & 10919 &  29 & S &*     	\nl
           16016+1506 &  16 03 55.81 & 	14 57 24.34 & 14.7 & 10924 &  26 & S &*     	\nl
           16016+1502 &  16 03 56.22 & 	14 53 39.95 & 15.9 & 10681 &  28 & E &*     	\nl
       A2151:[D80]016 &  16 03 56.59 & 	17 18 18.58 & 15.4 & 10023 &  33 & E &(23) \nl 
           16016+1630 &  16 03 57.13 & 	16 21 42.59 & 14.5 & 11592 & 100 & E &(13)  \nl 
           16016+1608 &  16 03 57.96 & 	16 00 01.62 & 13.8 &  9577 & 100 & S &(13)  \nl 
            1601+1741 &  16 04 00.30 & 	17 33 15.80 & 16.5 & 33709 &  50 & E &(23) \nl 
               sw-103 &  16 04 00.56 & 	17 15 12.81 & 19.0 & 11021 & 100 & S &(31)  \nl 
           16018+1725 &  16 04 02.71 & 	17 16 55.95 & 14.2 &  9908 & 100 & E &(13)  \nl 
            1601+1743 &  16 04 07.28 & 	17 34 54.26 & 15.7 & 11548 &  50 & S &(23) \nl 
       A2151:[D80]005 &  16 04 10.12 & 	17 12 24.91 & 16.4 & 10446 & 100 & S &(31)  \nl 
            1601+1807 &  16 04 10.13 & 	17 58 52.46 & 15.4 & 11616 &  50 & S &(23) \nl
           16019+1613 &  16 04 10.41 & 	16 05 15.18 & 14.7 & 10397 &  30 & S &*     	\nl
           16019+1618 &  16 04 13.22 & 	16 09 49.39 & 15.3 & 11826 &  35 & S &*     	\nl
           16020+1636 &  16 04 15.89 & 	16 27 36.36 & 15.0 & 12786 &  38 & S &*     	\nl
           16020+1534 &  16 04 16.19 & 	15 25 31.22 & 15.1 &  9225 &  25 & S &*     	\nl
           16020+1614 &  16 04 16.76 & 	16 05 41.61 & 15.1 & 13464 &  24 & S &*     	\nl
           16020+1614 &  16 04 16.93 & 	16 05 40.56 & 15.1 & 13549 &  24 & S &*     	\nl
               ne-398 &  16 04 18.03 & 	18 14 06.36 & 19.5 & 10602 & 100 & S &(31)  \nl 
            1602+1719 &  16 04 19.52 & 	17 10 51.02 & 14.6 & 10217 &  50 & E &(23) \nl 
       A2151:[D80]025 &  16 04 20.14 & 	17 26 10.07 & 15.5 & 10805 & 100 & S &(23) \nl 
           16021+1648 &  16 04 20.23 & 	16 39 48.63 & 15.2 & 13591 &  42 & E &*     	\nl
          H186=798932 &  16 04 21.05 & 	18 08 24.54 & 15.9 & 10868 & 100 & S &(31)  \nl 
           16021+1647 &  16 04 22.33 & 	16 38 30.98 & 15.1 & 13870 &  35 & E &*     	\nl
            1602+1800 &  16 04 22.94 & 	17 52 41.27 & 14.9 & 11170 &  50 & E &(23) \nl 
           16021+1650 &  16 04 23.01 & 	16 41 52.01 & 15.2 &  9366 & 100 & S &(24)  \nl 
           16022+2009 &  16 04 26.65 & 	20 00 32.40 & 15.4 & 10024 &  43 & S &*     	\nl
               N6040N &  16 04 26.69 & 	17 44 54.99 & 14.0 & 12404 & 100 & S &(13)  \nl 
               N6040S &  16 04 26.69 & 	17 44 30.01 & 14.0 & 12612 & 100 & E &(13)  \nl 
           16021+1455 &  16 04 28.20 & 	14 46 53.00 & 13.9 &  4702 &  49 & S &(16) \nl 
            1602+1747 &  16 04 28.77 & 	17 38 54.64 & 15.6 & 11354 &  50 & E &(23) \nl 
           16022+1736 &  16 04 30.72 & 	17 28 07.00 & 14.7 & 11987 & 100 & S &(24)  \nl 
           16022+1457 &  16 04 31.62 & 	14 49 07.72 & 13.9 &  4616 &  34 & E &* \nl 
                I1170 &  16 04 31.80 & 	17 43 16.82 & 15.4 &  9587 & 100 & E &(13)  \nl 
           16023+1649 &  16 04 32.38 & 	16 40 30.22 & 15.2 &  9622 &  35 & S &*     	\nl
          16023+1637S &  16 04 34.39 & 	16 28 51.93 & 15.4 & 11795 & 100 & S &(13)  \nl 
          16023+1637N &  16 04 34.39 & 	16 28 51.93 & 14.0 & 13527 & 100 & S &(13)  \nl 
            1602+1827 &  16 04 34.50 & 	18 18 54.00 & 16.1 & 10840 &  50 & E &(23) \nl 
               N6041B &  16 04 35.00 & 	17 43 00.01 & 20.2 & 11248 & 100 & E &(13)  \nl 
           16023+1701 &  16 04 35.61 & 	16 54 10.04 & 13.7 &  9222 & 100 & E &(13)  \nl 
               N6041A &  16 04 36.01 & 	17 43 45.01 & 12.4 & 10272 & 100 & E &(13)  \nl 
           16024+1758 &  16 04 37.80 & 	17 50 09.31 & 14.9 & 21267 &  42 & S &*     	\nl
                N6042 &  16 04 39.61 & 	17 42 02.30 & 13.8 & 10430 & 100 & E &(13)  \nl 
            1602+1810 &  16 04 40.04 & 	18 01 56.35 & 15.7 & 11259 &  50 & S &(23) \nl 
           16024+1503 &  16 04 40.80 & 	14 54 50.36 & 15.0 & 10874 &  31 & S &*     	\nl
           16024+1634 &  16 04 41.05 & 	16 25 47.35 & 14.2 &  9989 & 100 & E &(13)  \nl 
           16024+1647 &  16 04 41.42 & 	16 38 59.10 & 15.2 & 13581 &  52 & E &*     	\nl
           16024+1639 &  16 04 42.05 & 	16 30 40.93 & 15.5 & 10855 &  36 & E &*     	\nl
      A2151:[BO85]064 &  16 04 42.24 & 	17 41 00.46 & 15.8 & 10430 & 100 & S &(30)  \nl 
       A2151:[D80]051 &  16 04 42.78 & 	17 38 19.50 & 15.3 & 10112 & 100 & S &(30)  \nl 
           16025+1639 &  16 04 43.70 & 	16 31 20.86 & 14.8 & 13605 &  22 & E &*     	\nl
           16025+1735 &  16 04 45.30 & 	17 26 51.68 & 14.6 & 10413 & 100 & E &(24)  \nl 
            1602+1728 &  16 04 47.50 & 	17 20 50.82 & 15.3 & 10610 & 100 & S &(25)  \nl 
           16025+1701 &  16 04 48.43 & 	16 53 01.86 & 14.5 & 12553 & 100 & S &(13)  \nl 
            1602+1746 &  16 04 49.08 & 	17 38 38.90 & 16.1 & 11462 &  50 & E &(23) \nl 
           16025+1643 &  16 04 49.98 & 	16 35 01.00 & 14.2 &  9347 &  71 & S &(21)  \nl 
           16025+1552 &  16 04 51.71 & 	15 43 23.01 & 13.6 & 10670 & 100 & E &(13)  \nl 
           16027+1642 &  16 04 57.01 & 	16 34 06.60 & 15.4 & 13415 &  33 & S &*     	\nl
            1602+1819 &  16 04 58.40 & 	18 11 15.00 & 15.8 & 11355 &  20 & S &(12) \nl 
           16027+1657 &  16 04 58.62 & 	16 48 34.42 & 14.8 & 12957 &  27 & S &*     	\nl
                N6044 &  16 04 59.81 & 	17 52 11.60 & 13.9 &  9936 &  50 & E &(24)  \nl 
               N6043E &  16 05 01.50 & 	17 46 30.00 & 14.0 &  9798 & 100 & E &(13)  \nl 
       A2151:[D80]092 &  16 05 01.86 & 	17 49 50.73 & 15.4 & 10796 &  45 & S &(23) \nl 
   DKP160249.30+17570 &  16 05 04.16 & 	17 49 02.89 & 15.4 & 10433 & 200 & S &(12) \nl 
           16028+1641 &  16 05 04.54 & 	16 32 43.26 & 14.8 & 13649 &  36 & S &*     	\nl
               ce-200 &  16 05 06.66 & 	17 47 00.42 & 18.2 &  9927 & 100 & S &(31)  \nl 
           16028+1644 &  16 05 07.01 & 	16 35 45.17 & 14.9 & 14011 &  33 & S &*     	\nl
            1602+1747 &  16 05 07.05 & 	17 38 56.08 & 15.8 & 10186 & 100 & S &(23) \nl 
           16029+1642 &  16 05 07.28 & 	16 34 13.22 & 18.4 & 13334 &  39 & E &*     	\nl
                N6045 &  16 05 07.84 & 	17 45 27.11 & 14.0 &  9913 &  41 & S &(1)  \nl 
      A2151:[BO85]137 &  16 05 08.09 & 	17 48 55.11 & 17.1 & 13683 & 100 & S &(12) \nl 
                N6047 &  16 05 08.95 & 	17 43 47.17 & 13.7 &  9470 &  50 & E &(1)  \nl 
      A2151:[BO85]119 &  16 05 09.90 & 	17 51 20.16 & 17.1 &  9905 & 100 & S &(31)  \nl 
                I1173 &  16 05 12.70 & 	17 25 22.40 & 14.0 & 10871 & 100 & S &(24)  \nl 
      A2151:[BO85]053 &  16 05 14.57 & 	17 48 02.70 & 15.9 &  4923 & 100 & S &(30)  \nl 
           16030+1604 &  16 05 15.18 & 	15 55 32.38 & 14.2 & 13361 &  58 & S &*     	\nl
            1603+1740 &  16 05 15.36 & 	17 32 22.60 & 15.0 & 12299 &  50 & S &(23) \nl 
       A2151:[D80]061 &  16 05 15.80 & 	17 42 29.62 & 15.5 &  9432 &  45 & E &(23) \nl 
            1603+1830 &  16 05 16.01 & 	18 21 59.00 & 15.9 & 10926 &  50 & S &(23) \nl 
           16030+1518 &  16 05 18.44 & 	15 09 59.18 & 16.2 &  4802 &  24 & E &*     	\nl
            1603+1747 &  16 05 18.64 & 	17 39 19.80 & 16.5 &  9493 &  50 & E &(23) \nl 
       A2151:[D80]091 &  16 05 20.59 & 	17 52 01.74 & 17.7 & 11587 & 100 & S &(31)  \nl 
                N6050 &  16 05 20.90 & 	17 45 54.94 & 14.5 &  9511 & 100 & S &(24)  \nl 
                I1179 &  16 05 20.90 & 	17 45 54.94 & 16.3 & 11049 & 100 & S &(24)  \nl 
           16031+1552 &  16 05 21.10 & 	15 44 15.40 & 15.3 & 10652 &  22 & E &*     	\nl
           16031+1614 &  16 05 21.14 & 	16 06 15.41 & 15.3 & 10932 &  27 & S &*     	\nl
           16031+1607 &  16 05 21.34 & 	15 58 37.71 & 15.4 & 13055 &  30 & E &*     	\nl
           16031+1633 &  16 05 21.98 & 	16 24 36.47 & 15.2 & 13329 &  42 & E &*     	\nl
       A2151:[D80]102 &  16 05 21.98 & 	17 58 13.04 & 15.8 & 11851 &  45 & E &(23) \nl 
            1603+1759 &  16 05 22.59 & 	17 51 17.06 & 14.9 & 10834 &  50 & S &(23) \nl 
           16031+1620 &  16 05 22.67 & 	16 11 53.01 & 14.5 & 10088 &  71 & E &(21)  \nl 
          CGCG108-115 &  16 05 22.99 & 	14 38 49.99 & 13.9 & 14117 &  69 & S &(2)  \nl 
            1603+1816 &  16 05 23.50 & 	18 08 31.13 & 16.4 & 11220 & 100 & E &(24)  \nl 
           16031+1446 &  16 05 23.78 & 	14 38 52.01 & 13.9 & 12228 &  71 & S &(21)  \nl 
           16031+1619 &  16 05 24.11 & 	16 10 27.44 & 14.5 & 14003 &  32 & S &*     	\nl
           16031+1531 &  16 05 24.52 & 	15 22 32.99 & 15.8 & 12617 &  25 & S &*     	\nl
           16032+1751 &  16 05 26.38 & 	17 41 49.30 & 15.7 & 11070 & 100 & S &(27) \nl 
          16032+1635N &  16 05 26.49 & 	16 26 30.01 & 14.7 & 13598 & 100 & S &(13)  \nl 
        MCG+03-41-096 &  16 05 26.55 & 	17 54 36.32 & 14.4 & 12308 &  37 & S &(23) \nl 
                I1174 &  16 05 26.66 & 	15 01 35.22 & 13.0 &  4706 &  26 & S &(16) \nl 
                I1181 &  16 05 27.10 & 	17 35 55.35 & 14.0 & 10252 &  71 & E &(21)  \nl 
       A2151:[D80]090 &  16 05 27.63 & 	17 49 48.36 & 13.8 & 10253 &  45 & S &(23) \nl 
            1603+1828 &  16 05 27.67 & 	18 20 26.41 & 15.4 & 11550 &  50 & S &(23) \nl 
          16032+1635S &  16 05 29.29 & 	16 26 04.99 & 13.9 & 13271 & 100 & S &(13)  \nl 
       A2151:[D80]097 &  16 05 29.33 & 	17 55 42.49 & 15.1 & 11905 &  45 & S &(23) \nl 
           A1603+1748 &  16 05 29.69 & 	17 40 50.52 & 14.7 & 11086 &  20 & E &(12) \nl 
           16033+1933 &  16 05 29.74 & 	19 24 34.09 & 15.2 & 28680 &  57 & E &*     	\nl
           16032+1530 &  16 05 30.17 & 	15 22 25.18 & 15.6 & 10558 &  28 & E &*     	\nl
           16033+2010 &  16 05 30.44 & 	20 01 40.26 & 15.4 &  9293 &  36 & E &*     	\nl
                N6054 &  16 05 30.92 & 	17 46 13.58 & 14.8 & 11177 &  40 & S &(20) \nl 
                N6056 &  16 05 31.20 & 	17 57 48.60 & 13.4 & 11701 & 138 & E &(21)  \nl 
                N6055 &  16 05 32.57 & 	18 09 33.70 & 13.5 & 11315 & 100 & E &(21)  \nl 
                I1178 &  16 05 33.00 & 	17 36 05.01 & 12.9 & 10200 &  71 & E &(21)  \nl 
           16033+1630 &  16 05 33.35 & 	16 22 00.30 & 16.0 & 13516 &  30 & E &*     	\nl
           16033+1744 &  16 05 33.82 & 	17 35 35.96 & 14.7 & 10192 &  26 & S &*     	\nl
           16033+1640 &  16 05 35.81 & 	16 31 33.46 & 14.8 &  4630 &  36 & S &*     	\nl
           16033+1825 &  16 05 36.52 & 	18 16 21.97 & 14.2 & 11285 & 100 & S &(13)  \nl 
                I1182 &  16 05 36.74 & 	17 48 06.95 & 13.9 & 10091 &  40 & E &(20) \nl 
                I1183 &  16 05 38.19 & 	17 46 00.01 & 14.3 & 10038 &  75 & E &(1)  \nl 
           16034+1630 &  16 05 38.76 & 	16 22 22.55 & 15.4 & 13406 &  42 & E &*     	\nl
           16034+1634 &  16 05 39.12 & 	16 26 09.89 & 16.0 & 40569 &  50 & S &*     	\nl
                N6057 &  16 05 39.59 & 	18 09 50.01 & 14.2 & 10443 & 100 & E &(13)  \nl 
           1603+1756A &  16 05 39.88 & 	17 48 02.16 & 13.9 &  9973 & 100 & E &(24)  \nl 
                N6053 &  16 05 40.20 & 	18 03 16.99 & 15.0 & 11993 & 100 & E &(13)  \nl 
           16034+1634 &  16 05 40.37 & 	16 25 45.05 & 15.9 & 40031 &  52 & S &*     	\nl
           16034+1814 &  16 05 40.79 & 	18 06 26.14 & 15.2 & 11401 &  30 & E &*     	\nl
           1603+1756B &  16 05 40.85 & 	17 48 02.23 & 13.9 &  9979 & 100 & S &(24)  \nl 
           16034+1456 &  16 05 40.95 & 	14 47 56.65 & 14.8 & 12218 &  28 & S &*     	\nl
           16035+1708 &  16 05 43.42 & 	16 59 34.37 & 15.2 & 14048 &  29 & E &*     	\nl
           16035+1624 &  16 05 43.94 & 	16 15 34.27 & 14.9 & 13093 &  26 & S &*     	\nl
                I1186 &  16 05 44.34 & 	17 21 44.43 & 13.7 & 11043 & 100 & S &(24)  \nl 
          16035+1620N &  16 05 44.70 & 	16 12 10.01 & 14.7 & 13079 &  20 & S &(7)  \nl 
                I1185 &  16 05 44.95 & 	17 42 56.49 & 13.7 & 10297 & 100 & S &(13)  \nl 
            1603+1742 &  16 05 45.42 & 	17 34 55.53 & 15.0 & 12206 & 100 & S &(25)  \nl 
            1603+1726 &  16 05 46.32 & 	17 18 19.55 & 15.7 & 11088 & 100 & S &(5) \nl 
          16035+1620S &  16 05 46.32 & 	16 11 43.01 & 16.0 & 12079 &  20 & E &(7)  \nl 
           16035+1809 &  16 05 46.32 & 	18 01 00.98 & 14.4 & 12212 & 100 & S &(13)  \nl 
           16035+1555 &  16 05 47.08 & 	15 47 26.56 & 17.7 & 12450 & 100 & S &(13)  \nl 
               ne-264 &  16 05 47.49 & 	18 23 02.26 & 17.2 & 12044 & 100 & S &(31)  \nl 
           16036+1840 &  16 05 49.49 & 	18 32 04.99 & 14.5 & 11139 & 100 & E &(13)  \nl 
            1603+1736 &  16 05 50.03 & 	17 28 45.80 & 15.9 & 10255 &  50 & E &(23) \nl 
           16035+1510 &  16 05 50.71 & 	15 01 51.96 & 15.0 & 11880 &  26 & S &*     	\nl
                 H186 &  16 05 52.22 & 	18 27 57.56 & 16.2 & 11645 & 100 & S &(31)  \nl 
           16036+1821 &  16 05 52.26 & 	18 13 13.98 & 14.4 & 10739 &  20 & S &(6)  \nl 
           16036+1620 &  16 05 52.76 & 	16 11 59.14 & 14.8 & 13351 & 160 & S &*     	\nl
       A2151:[D80]013 &  16 05 53.09 & 	17 18 28.01 & 16.0 &  9728 &  45 & E &(23) \nl 
           16036+1551 &  16 05 53.66 & 	15 42 30.53 & 14.7 & 13148 &  34 & S &*     	\nl
      IRASF16035+1728 &  16 05 53.88 & 	17 20 26.05 & 15.8 & 10265 & 100 & S &(23) \nl 
            1603+1826 &  16 05 54.78 & 	18 18 42.98 & 15.8 & 11622 &  50 & S &(23) \nl 
               ce-048 &  16 05 55.72 & 	17 42 38.56 & 20.4 & 11145 & 100 & S &(31)  \nl 
            1603+1749 &  16 05 56.68 & 	17 41 30.99 & 16.0 & 11540 &  50 & E &(23) \nl 
           16037+1552 &  16 05 57.55 & 	15 43 31.94 & 15.0 & 13176 &  35 & S &*     	\nl
               ne-240 &  16 05 58.31 & 	18 24 40.97 & 17.6 & 11645 & 100 & S &(31)  \nl 
           16037+1554 &  16 05 58.45 & 	15 45 39.49 & 14.6 & 12154 &  27 & S &*     	\nl
        160343+164242 &  16 06 00.09 &  16 34 38.39 & 14.8 & 12849 &  30 & S &* \nl
            1603+1753 &  16 06 00.18 & 	17 45 52.45 & 18.6 & 11959 & 100 & S &(25)  \nl 
           16037+1643 &  16 06 00.26 & 	16 34 38.39 & 14.8 & 12230 & 190 & E &*     	\nl
       A2151:[D80]112 &  16 06 00.43 & 	18 04 54.48 & 15.5 & 11272 &  45 & E &(23) \nl 
           16038+1820 &  16 06 00.50 & 	18 11 43.01 & 13.8 & 11215 & 100 & S &(13)  \nl 
            1603+1815 &  16 06 01.91 & 	18 06 42.01 & 14.9 & 10967 &  50 & S &(23) \nl 
           16038+1849 &  16 06 02.05 & 	18 40 11.60 & 14.4 & 11751 & 100 & S &(13)  \nl 
            1603+1815 &  16 06 02.49 & 	18 06 57.60 & 16.0 & 11030 & 100 & S &(23) \nl 
            1603+1750 &  16 06 03.06 & 	17 42 05.65 & 16.2 & 10829 &  50 & E &(23) \nl 
        160346+161835 &  16 06 03.45 &  16 10 33.20 & 14.9 & 10414 &  40 & E &*  \nl
            1603+1810 &  16 06 05.58 & 	18 02 09.82 & 15.2 & 12190 & 100 & E &(25)  \nl 
               ne-208 &  16 06 05.67 & 	18 16 43.43 & 18.7 & 11556 & 100 & S &(31)  \nl 
             MGT95:14 &  16 06 05.93 & 	18 09 20.45 & 19.8 & 11467 & 100 & S &(31)  \nl 
           16039+1845 &  16 06 06.19 & 	18 36 25.99 & 14.2 & 11330 & 100 & S &(13)  \nl 
            1603+1735 &  16 06 07.24 & 	17 27 38.91 & 14.8 & 10699 &  50 & S &(23) \nl 
           A1603+1800 &  16 06 08.75 & 	17 52 34.00 & 16.9 & 10762 & 197 & S &(22)  \nl 
           16039+1619 &  16 06 10.12 & 	16 11 00.46 & 15.1 & 12766 &  24 & S &*     	\nl
       A2151:[D80]140 &  16 06 11.85 & 	18 19 43.21 & 15.9 & 10939 & 100 & S &(23) \nl 
            1603+1726 &  16 06 11.99 & 	17 18 06.19 & 14.7 & 10128 &  50 & S &(23) \nl 
            1603+1812 &  16 06 13.71 & 	18 04 40.33 & 15.0 & 11002 &  50 & E &(23) \nl 
           16040+2016 &  16 06 13.77 & 	20 08 06.32 & 15.5 & 51910 &  62 & S &*     	\nl
           16040+1627 &  16 06 13.84 & 	16 19 22.77 & 14.9 & 13215 &  31 & S &*     	\nl
            1603+1805 &  16 06 13.86 & 	17 57 15.34 & 15.2 & 11440 &  32 & S &(22)  \nl 
           16040+1833 &  16 06 14.15 & 	18 24 58.36 & 14.3 & 11161 & 100 & S &(24)  \nl 
                I1189 &  16 06 14.61 & 	18 10 55.42 & 14.3 & 11810 &  40 & S &(20) \nl 
                N6061 &  16 06 16.02 & 	18 14 59.50 & 13.9 & 11305 & 100 & E &(13)  \nl 
           16040+1848 &  16 06 16.14 & 	18 39 52.92 & 15.8 & 10920 &  54 & S &*     	\nl
           16040+1610 &  16 06 16.45 & 	16 02 20.94 & 14.9 & 12953 &  33 & E &*     	\nl
           16040+1627 &  16 06 16.67 & 	16 18 46.69 & 15.3 & 13470 &  33 & S &*     	\nl
           A1604+1829 &  16 06 17.79 & 	18 21 42.01 & 15.1 & 11099 &  63 & S &(12) \nl 
           16040+1552 &  16 06 18.23 & 	15 43 37.45 & 15.3 & 10904 &  38 & S &*     	\nl
           16040+1640 &  16 06 19.22 & 	16 32 18.63 & 14.2 & 10785 & 100 & E &(13)  \nl 
           16040+1634 &  16 06 19.47 & 	16 25 52.64 & 13.9 & 10130 & 100 & S &(13)  \nl 
       A2151:[D80]076 &  16 06 20.74 & 	17 47 15.79 & 16.0 & 11950 &  45 & E &(23) \nl 
               ne-142 &  16 06 22.48 & 	18 00 02.70 & 19.6 & 11711 & 100 & S &(31)  \nl 
            1604+1803 &  16 06 22.50 & 	17 55 43.89 & 13.8 & 10494 &  50 & S &(23) \nl 
           A1604+1803 &  16 06 22.68 & 	17 55 40.91 & 15.3 & 26579 & 218 & S &(22)  \nl 
          16041+1549S &  16 06 25.70 & 	15 41 04.99 & 13.9 & 13201 &  23 & E &(17)  \nl 
          16041+1549N &  16 06 25.99 & 	15 41 31.99 & 14.5 & 12044 &  22 & S &(17)  \nl 
            1604+1824 &  16 06 29.09 & 	18 16 07.32 & 15.3 & 11632 &  50 & S &(23) \nl 
            1604+1824 &  16 06 29.99 & 	18 16 05.38 & 16.5 & 11998 &  50 & S &(23) \nl 
           16042+1544 &  16 06 30.13 & 	15 36 12.89 & 14.3 & 12670 &  51 & S &*     	\nl
                I1193 &  16 06 32.11 & 	17 42 50.51 & 13.9 & 12034 &  30 & S &(18)\nl 
                I1192 &  16 06 33.15 & 	17 46 33.56 & 14.2 & 11500 &  46 & S &(18) \nl 
            1604+1725 &  16 06 34.31 & 	17 17 49.63 & 14.9 & 21295 &  50 & S &(23) \nl 
           16043+1801 &  16 06 35.21 & 	17 53 32.71 & 14.7 & 11049 & 100 & S &(13)  \nl 
       A2151:[D80]055 &  16 06 35.93 & 	17 43 21.75 & 15.2 & 10993 & 100 & E &(30)  \nl 
       A2151:[D80]054 &  16 06 35.94 & 	17 41 45.74 & 15.5 & 11967 & 100 & S &(30)  \nl 
               ne-112 &  16 06 37.61 & 	18 23 48.87 & 16.7 & 11046 & 100 & S &(31)  \nl 
           16044+1550 &  16 06 38.29 &  15 41 51.47 & 14.7 & 13017 &  28 & E &*\nl
       A2151:[D80]073 &  16 06 38.84 & 	17 47 00.93 & 14.7 & 11190 & 100 & E &(30)  \nl 
           16044+1704 &  16 06 39.52 & 	16 56 14.28 & 14.9 & 12092 &  30 & S &*     	\nl
                I1194 &  16 06 39.60 & 	17 45 38.01 & 14.1 & 11642 &  65 & E &(1)  \nl 
    A2151:[HKT95]4028 &  16 06 40.07 & 	17 35 18.99 & 16.7 & 14099 & 241 & S &(22)  \nl 
                I1195 &  16 06 40.79 & 	17 11 30.05 & 14.1 & 12121 & 100 & S &(13)  \nl 
            1604+1743 &  16 06 41.04 & 	17 35 31.06 & 15.1 & 12226 &  51 & S &(18) \nl 
          A1604+1557B &  16 06 41.04 & 	15 49 18.99 & 17.9 & 11335 &  74 & E &(22)  \nl 
          A1604+1557A &  16 06 41.04 & 	15 50 17.02 & 16.8 & 11437 &  80 & S &(22)  \nl 
           16044+1627 &  16 06 42.09 & 	16 19 11.10 & 13.3 & 11015 &  26 & S &(19) \nl 
             BO85:130 &  16 06 44.51 & 	18 14 47.40 & 17.3 & 11600 & 100 & S &(31)  \nl 
       A2151:[D80]071 &  16 06 47.92 & 	17 47 22.53 & 16.9 & 12561 &  45 & E &(23) \nl 
            1604+1746 &  16 06 48.20 & 	17 38 53.52 & 15.1 & 11201 &  50 & S &(23) \nl 
           16045+1738 &  16 06 48.46 & 	17 29 34.55 & 14.6 & 11199 &  20 & E &(12) \nl 
           16045+1456 &  16 06 50.40 & 	14 47 46.00 & 14.2 & 11373 &  29 & S &* \nl 
            1604+1821 &  16 06 53.24 & 	18 13 21.86 & 15.0 & 10845 &  50 & S &(23) \nl 
            1604+1759 &  16 06 56.95 & 	17 51 23.11 & 15.9 & 11517 &  20 & E &(12) \nl 
           16047+1618 &  16 06 58.21 &  16 09 43.49 & 14.5 & 12221 &  32 & E &* \nl
            1604+1809 &  16 07 01.63 & 	18 01 20.39 & 14.4 & 10870 &  50 & S &(23) \nl 
           16047+1544 &  16 07 03.18 & 	15 35 35.41 & 14.2 & 11800 &  20 & S &(6)  \nl 
            1604+1758 &  16 07 06.06 & 	17 50 56.68 & 14.9 & 11582 &  50 & E &(23) \nl 
           16049+1846 &  16 07 09.87 & 	18 38 28.89 & 13.8 & 11599 &  45 & S &*     	\nl
           16049+1643 &  16 07 10.53 & 	16 34 32.41 & 14.9 & 11463 &  46 & S &*     	\nl
           16050+1835 &  16 07 11.86 & 	18 27 11.16 & 15.2 & 10847 &  28 & S &*     	\nl
            1605+1749 &  16 07 17.00 & 	17 41 46.36 & 15.4 & 10977 &  50 & S &(12) \nl 
            1605+1756 &  16 07 18.77 & 	17 48 06.48 & 15.7 & 27047 &  50 & E &(1)  \nl 
           16050+1559 &  16 07 19.43 &  15 50 58.92 & 14.7 & 11913 &  32 & S &* \nl
            1605+1747 &  16 07 24.13 & 	17 39 41.83 & 14.6 & 11546 &  20 & S &(12) \nl 
            1605+1822 &  16 07 25.46 & 	18 14 39.95 & 15.9 & 10634 &  50 & E &(23) \nl 
           16052+1509 &  16 07 28.85 & 	15 00 40.61 & 14.9 & 11241 &  38 & S &*     	\nl
            1605+1810 &  16 07 31.51 & 	18 02 27.31 & 14.8 & 11078 &  50 & S &(23) \nl 
            1605+1836 &  16 07 38.21 & 	18 28 45.99 & 14.4 & 11169 &  22 & S &(26) \nl 
            1605+1825 &  16 07 54.30 & 	18 17 43.80 & 15.5 & 10810 &  20 & S &(12) \nl 
          CGCG108-157 &  16 08 10.72 & 	16 46 17.80 & 14.3 & 11569 &  39 & S &(28)\nl 
          A1605+1625B &  16 08 12.44 & 	16 17 41.89 & 16.6 & 76003 &  53 & S &(22)  \nl 
          A1605+1625A &  16 08 13.42 & 	16 18 07.99 & 15.5 & 12351 & 100 & E &(22)  \nl 
           16060+1842 &  16 08 14.89 & 	18 34 15.56 & 14.8 & 10575 &  28 & S &*     	\nl
           16060+1732 &  16 08 17.51 &  17 24 35.93 & 15.1 & 10363 &  31 & E &*  \nl
           16062+1659 &  16 08 27.96 &  16 51 23.90 & 15.1 & 11781 &  46 & E &* \nl 
           16062+1821 &  16 08 29.26 & 	18 12 43.13 & 14.5 & 11008 &  26 & E &*     	\nl
           A1606+1637 &  16 08 42.79 & 	16 30 11.99 & 15.1 & 12243 & 401 & E &(22)  \nl 
           16065+1654 &  16 08 45.20 & 	16 45 33.01 & 13.9 & 10596 & 100 & S &(25)  \nl 
           A1606+1737 &  16 08 59.06 & 	17 29 56.90 & 15.7 & 10786 & 200 & E &(22)  \nl 
          A1606+1612B &  16 09 07.63 & 	16 04 21.43 & 18.1 & 12855 & 355 & E &(22)  \nl 
          A1606+1612A &  16 09 07.63 & 	16 04 21.43 & 15.2 & 22475 & 100 & S &(22)  \nl 
           A1606+1547 &  16 09 12.10 & 	15 39 57.71 & 17.3 & 13271 & 326 & E &(22)  \nl 
            1607+1919 &  16 09 17.10 & 	19 12 07.16 & 16.4 & 10762 & 100 & E &(14)  \nl 
          A1607+1637A &  16 09 28.19 & 	16 29 28.75 & 16.5 & 12315 & 167 & E &(22)  \nl 
          A1607+1637B &  16 09 28.66 & 	16 29 36.78 & 20.0 & 13319 & 167 & E &(22)  \nl 
           16073+1823 &  16 09 31.57 & 	18 15 10.08 & 14.5 &  6373 &  81 & S &*     	\nl
           16074+1854 &  16 09 39.38 & 	18 45 57.03 & 14.6 & 10737 &  22 & S &*     	\nl
           16076+1749 &  16 09 49.70 &  17 40 43.22 & 14.8 & 10162 &  21 & E &*  \nl
           16075+1544 &  16 09 49.97 & 	15 35 54.85 & 14.9 & 13814 &  26 & E &*     	\nl
                N6073 &  16 10 10.92 & 	16 41 54.53 & 15.1 &  4590 &	8 & S &(11) \nl 
           A1607+1512 &  16 10 10.99 & 	15 05 30.01 & 15.9 & 17606 & 158 & E &(22)  \nl 
           16080+1545 &  16 10 17.62 & 	15 37 14.01 & 14.6 & 13405 &  17 & E &* \nl 
           16084+1911 &  16 10 38.68 &  19 03 26.75 & 14.8 & 10855 &  39 & S &* \nl  
           A1608+1505 &  16 10 43.83 & 	14 57 51.59 & 17.4 & 13577 & 233 & S &(22)  \nl 
           16086+1914 &  16 10 49.72 & 	19 05 51.25 & 14.5 & 11005 &  22 & S &*     	\nl
           16086+1806 &  16 10 50.56 & 	17 58 30.15 & 14.0 & 10967 &  20 & S &(16) \nl 
           16086+1711 &  16 10 51.31 & 	17 03 20.99 & 13.8 & 10184 &  14 & S &(24)  \nl 
           16088+1838 &  16 11 05.28 & 	18 29 54.99 & 14.4 & 10837 & 100 & S &* \nl 
           A1609+1528 &  16 11 26.38 & 	15 20 30.33 & 16.9 & 14234 & 100 & S &(22)  \nl 
            1609+1707 &  16 11 33.54 & 	16 59 18.89 & 14.7 & 26949 &  47 & S &* \nl 
            1609+1700 &  16 11 47.65 & 	16 52 25.79 & 15.3 &  4569 &  41 & S &* \nl 
           16097+1654 &  16 11 56.83 & 	16 46 40.04 & 14.3 & 10246 &  34 & S &*     	\nl
           16099+1446 &  16 12 09.85 &  14 38 14.03 & 15.1 &  9611 &  20 & S &*  \nl
           16102+1619 &  16 12 30.53 & 	16 11 01.57 & 14.8 & 10457 &  26 & E &*     	\nl
           16106+1701 &  16 12 50.94 & 	16 53 08.99 & 15.2 & 10253 &  54 & E &*     	\nl
\tablebreak
           16109+1828 &  16 13 11.66 &  18 20 50.75 & 15.0 & 10827 &  35 & E &* \nl
           16110+1720 &  16 13 15.20 & 	17 12 25.49 & 17.4 &  1089 &  25 & S &(29) \nl 
           16116+1503 &  16 13 54.77 & 	14 55 35.80 & 14.0 & 14036 &  29 & S &*     	\nl
           16120+1926 &  16 14 11.26 & 	19 18 54.18 & 14.4 &  9213 &  57 & S &* \nl
\enddata
\tablerefs{
(*) new velocities, this paper;
(1) de Vaucouleurs et al.\ 1976; (2) de Vaucouleurs et al.\ 1991; (3) Huchra \& Sargent 1973;
(4) Denisyuk et al.\ 1976; (5) Arakelyan et al.\ 1972; (6) Giovanelli \& Haynes 1981; 
(7) Giovanelli \& Haynes 1985; (8) Freudling et al. 1991; (9) Giovanardi \& Salpeter 1985
(10) Schneider et al.\ 1990; (11) Mould et al.\ 1993; (12) Bird et al.\ 1993;
(13) Tarenghi et al.\ 1979; (14) Lipovetskii \& Stepanyan 1986 (15) Ulrich 1976; 
(16) Huchra et al.\ 1983; (17) White et al.\ 1983; (18) Zabludoff et al.\ 1990; 
(19) Zabludoff et al.\ 1993a; (20) Schommer et al.\ 1981; (21) Rood 1981 (22) Hopp et al.\ 1995; 
(23) Dressler \& Shectman 1988; (24) Palumbo et al.\ 1983; (25) Huchtmeier \& Richter 1989;
(26) Strauss et al.\ 1992; (27) Lawrence et al.\ 1997; (28) Scodeggio \& Gavazzi 1993;
(29) Bothun et al.\ 1985; (30) Maccagni et al.\ 1995; (31) Dickey 1997
}
\end{deluxetable}

\begin{table}
\caption{Substructure statistics for supercluster and clusters\label{substruct}}
\begin{tabular}{lrccccc}
Name & $N$ & skewness & kurtosis & $\Delta$\tablenotemark{a} & $\alpha$ & $\epsilon$\nl
\tableline
supercluster & 414 & 1.68 &  4.77 & $99.1  $ & $94.4  $ & 0.026 \\
A2151 & 143 &-0.10 & -0.61 &  99.9  &  99.9  & 0.000 \\
A2152 &  56 &-0.54 & -0.65 &  99.9  &  99.9  & 0.6   \\
A2147 &  93 & 0.14 & -0.50 &  99.5  &  84.3  & 66.3  \\
\end{tabular}
\tablenotetext{a}{Values for $\Delta$,$\alpha$, and $\epsilon$ are percentage of 
Monte Carlo realizations with values of the statistics less than that for the cluster.}
\end{table}

\begin{table}
\caption{Dynamical parameters of clusters\label{cl-param}}
\begin{tabular}{ccccc}
 Cluster & $\bar{v}$  & ${\sigma}_v$ & $M_{VT}$ & $M_P$ \\
     & km~s$^{-1}$ & km~s$^{-1}$ & $h^{-1}M_{\sun}$ & $h^{-1}M_{\sun}$ \\
\tableline
A2151 & 11004 $\pm$  59 & 705$^{+46}_{-39}$   & $(7.0\pm 0.9)\times 10^{14}$ & $(8.1 \pm 1.0)\times 10^{14}$ \\ 
A2152 &	12942 $\pm$  97 & 715$^{+81}_{-61}$ & $(7.2\pm 1.7)\times 10^{14}$ & $(1.6 \pm 0.3)\times 10^{15}$ \\ 
A2147 &	10492 $\pm$ 85 & 821$^{+68}_{-55}$ & $(1.3\pm 0.2)\times 10^{15}$ & $(1.4 \pm 0.2)\times 10^{15}$ \\ 
dispersed & 11639 $\pm$ 128 & 1407$^{+100}_{-83}$ \\ 
\end{tabular}
\end{table}

\begin{table}
\caption{Other mass determinations for Hercules clusters\label{other-mass}}
\begin{tabular}{lcccc}
 Cluster & method  & $M$ ($h^{-1}M_{\sun}$) & Ref.\\
\tableline
A2151 & VT & $8.65 \times 10^{14}$ & (1) \\
A2151 & VT & $4.4  \times 10^{14}$ & (2) \\
A2151 & PM & $9.5  \times 10^{14}$ & (2) \\
A2151 & VT & $(1.07 \pm 0.21) \times 10^{15}$ & (3) \\
A2151 & X-ray ($<678$kpc) & $4.6  \times 10^{13}$ & (4) \\
A2152 & VT & $2.59 \times 10^{15}$ & (1) \\
A2147 & VT & $2.01 \times 10^{15}$ & (1) \\
A2147 & X-ray & $4.9^{+2.6}_{-1.0} \times 10^{15}$ & (5) \\
\end{tabular}
\tablerefs{
(1) Tarenghi et al.\ 1980; (2) Bird et al.\ 1993; (3) Escalera et al.\ 1994;
(4) Huang \& Sarazin 1996; (5) Henriksen \& White 1996 
}
\end{table}

\begin{table}
\caption{Luminosity parameters for clusters\label{lum-param}}
\begin{tabular}{lccc}
Cluster & $L_R $ & $(M/L)_{R,VT}$ & $(M/L)_{R,PM}$\\
  & $h^{-2} L_{\sun}$ & $(h (M/L)_{\sun}) $ & $(h (M/L)_{\sun}) $ \\
\tableline
A2151 & (1.9 $\pm$ 0.4) $\times 10^{12}$ &  374 $\pm$  92 & 430 $\pm$ 103  \\
A2152 & (1.3 $\pm$ 0.2) $\times 10^{12}$ &  542 $\pm$ 145 & 1206 $\pm$ 288  \\
A2147 & (1.6 $\pm$ 0.3) $\times 10^{12}$ &  774 $\pm$ 190 & 879 $\pm$ 204 \\
\end{tabular}
\end{table}

\begin{table}
\caption{Kinematical parameters of ellipticals and spirals for clusters\label{esvel}}
\begin{tabular}{lcccccc}
 Cluster & $N_E$ & ${\bar{v}}_E$ (km~s$^{-1})$ 
& ${\sigma}_E$(km~s$^{-1})$ & $N_S$
& ${\bar{v}}_S$(km~s$^{-1})$ & ${\sigma}_S$(km~s$^{-1})$\\
\tableline
A2151 & 56  &  10803$\pm$104  &  779$^{+87}_{-65}$ &   87 & 11133$\pm$68  &  641$^{+55}_{-44}$ \\   
A2152 & 20  &  13011$\pm$157  &  684$^{+150}_{-93}$ &  37 & 12905$\pm$126 &  769$^{+110}_{-77}$ \\   
A2147 & 43  &  10476$\pm$110  &  715$^{+94}_{-68}$ &   50 & 10505$\pm$129 &  906$^{+109}_{-80}$ \\   
A2151 SW & 7 & 10150$\pm$61   & 141$^{+78}_{-41}$ &    7 & 11111$\pm$129 &  593$^{+302}_{-123}$ \\   
\end{tabular}
\end{table}

\begin{table}
\caption{Distances to clusters in the Hercules supercluster\label{tfdata}}
\begin{tabular}{lclclcc}
Cluster & $\bar{v}$ & $N$\tablenotemark{a} & d (Mpc)\tablenotemark{a} 
&$N$\tablenotemark{b} & d(Mpc)\tablenotemark{b} & $cz/H_0$\tablenotemark{c} \\
\tableline
2151 & 11004 & 7 &  87 $\pm$ 11\tablenotemark{d} & 11 & 113 $\pm$ 12 & 89 $\pm$ 15\\
2152 & 12942 & 0 &  \nodata     &  2 & 119 $\pm$ 12 & 152 $\pm$ 26 \\
2147 & 10492 & 7 &  92 $\pm$ 11 &  2 &  76 $\pm$ 9 & 85 $\pm$ 14 \\
\end{tabular}
\tablenotetext{a}{Buta \& Corwin 1986, `GCH' data}
\tablenotetext{b}{Buta \& Corwin 1986, `other' data}
\tablenotetext{c}{Postman \& Lauer 1995}
\tablenotetext{d}{Errors calculated from BC's TF equation and quoted errors in observables.} 
\end{table}

\begin{table}
\caption{Binding ratios for pairs of clusters\label{2body}}
\begin{tabular}{lcc}
Pair & binding ratio & $\alpha$ range\\
\tableline
A2147-A2151 & 0.05   & $13\arcdeg <\alpha<88\arcdeg $  \\
A2151-A2152 & 0.88   &  not bound   \\
A2147-A2152 & 0.50   &  not bound  \\
A2147/51+2152 & 0.28 & $35\arcdeg <\alpha<73\arcdeg $  \\
\end{tabular}
\end{table}

\begin{table}
\caption{CCD Photometry for A2151 galaxies\label{phot-2151}}
\begin{tabular}{lllll}
 Name & $\alpha $ (J2000) & $\delta $ (J2000)  & $B$ & $B-R$\\
\tableline
16028+1756  	& 16 05 05 & 17 47 44 & 18.68  & 1.59    \\    
16028+1757  	& 16 05 05 & 17 48 25 & 18.61  & 1.38    \\    
NGC6045         & 16 05 08 & 17 45 27 & 15.15  & 1.56    \\    
A2151:[BO85]137 & 16 05 08 & 17 48 55 & 18.04  & 1.29    \\    	   
NGC6047   	& 16 05 09 & 17 43 47 & 15.28  & 1.62    \\    
A2151:[BO85]053 & 16 06 15 & 17 48 03 & 16.90  & 1.51    \\    	   
A2151:[D80]061  & 16 05 16 & 17 42 30 & 16.94  & 1.66    \\    	   
16030+1750  	& 16 05 18 & 17 41 36 & 19.47  & 1.62    \\    
16031+1755  	& 16 05 21 & 17 46 55 & 18.35  & 1.47    \\    
16032+1751      & 16 05 26 & 17 41 49 & 17.04  & 1.26    \\    	   
16032+1757  	& 16 05 28 & 17 48 57 & 18.19  & 1.41    \\    
16032+1757  	& 16 05 29 & 17 40 51 & 16.19  & 1.58    \\    
A1603+1748      & 16 05 30 & 17 46 14 & 15.92  & 1.21    \\    	   
NGC6054   	& 16 05 31 & 17 41 18 & 17.38  & 1.53    \\    
16033+1749      & 16 05 33 & 17 48 07 & 15.53  & 1.60    \\     
IC1182    	& 16 05 37 & 17 46 00 & 15.62  & 1.71    \\    
IC1183    	& 16 05 38 & 17 46 00 & 15.62  & 1.71    \\
\end{tabular}
\tablecomments{Some of these galaxies do not appear in Table 1 because
they are below our survey magnitude limit.}
\end{table}

\clearpage

\end{document}